\documentclass[journal=jpclcd,manuscript=letter]{achemso}
\usepackage{amsmath,amssymb}
\usepackage{float}

\usepackage[T1]{fontenc}
\usepackage[utf8]{luainputenc}
\usepackage{float}
\usepackage{multirow}
\usepackage{lipsum}
\usepackage{mathrsfs}
\usepackage{setspace}
\usepackage{droidsans}
\usepackage{balance}
\usepackage{times,mathptmx}
\usepackage{braket}
\usepackage{sectsty}
\usepackage{graphicx}
\usepackage{amsmath,amssymb}
\usepackage{mathrsfs}
\usepackage{multirow} 
\usepackage{lastpage}
\usepackage{adjustbox}
\usepackage{upgreek}
\usepackage{color}
\usepackage{array}
\usepackage{float}
\usepackage{fancyhdr}
\usepackage{fnpos}
\usepackage{adjustbox}
\usepackage[english]{babel}
\DeclareUnicodeCharacter{0308}{~}
\usepackage[version=3]{mhchem} 

\author{Megha Arya}
\email{meghaphy2@gmail.com [MA]}
\author{Preeti Bhumla}
\author{Sajjan Sheoran}
\author {Suraj}
\author{S. Bhattacharya}
\email{saswata@physics.iitd.ac.in[SB]}
\phone{+91-11-2659 1359}
\affiliation[Indian Institute of Technology Delhi]
{Department of Physics, Indian Institute of Technology Delhi, New Delhi, India}
\title[An \textsf{achemso} demo]
{Rashba and Dresselhaus effects in doped methylammonium lead halide perovskite MAPbI$_3$}
\keywords{DFT, Rashba, Dresselhaus, symmetry, spin-orbit coupling, spin texture, hybrid perovskites}
\begin{document}
\sloppy

\begin{abstract}
\noindent Inorganic-organic lead halide perovskites, particularly methylammonium lead halide (MAPbI$_3$) perovskite, is perceived to be a promising material for optoelectronics and spintronics. However, lead toxicity and instability under air and moisture restrict its practical uses. Hence, it is essential to reduce lead extent by substituting appropriate alternatives. Here, we substitute Sn and Ge in cubic MAPbI$_3$ and compare various properties of hybrid perovskites by employing state-of-the-art first-principles-based methodologies, viz., density functional theory (DFT) with semilocal and hybrid functional (HSE06) and generalized gradient approximation (PBE) combined with spin-orbit coupling (SOC). We mainly study the Rashba-Dresselhaus (RD) effect, which occurs here due to two major mechanisms breaking inversion symmetry, i.e., static and dynamic, and the presence of heavy elements contributing to significant SOC. We find non-negligible spin-splitting effects in the conduction band minimum (CBm) and valence band maximum (VBM) for hybrid perovskites. For a deeper understanding of the observed spin-splitting, the spin textures are analyzed and Rashba coefficients are calculated. We find that Dresselhaus effect comes into play in substituted hybrids in addition to the usual Rashba effect observed in pristine compound. We also observe that the strength of Rashba spin-splitting can be substantially tuned on application of uniaxial strain ($\pm5\%$). Also, we notice that some of the hybrids are mechanically stable and ductile. Hence these hybrid perovskites can prove to be potent for perovskite-based spintronic applications. 

  \begin{tocentry}
  \begin{figure}[H]
  	\includegraphics[width=5cm]{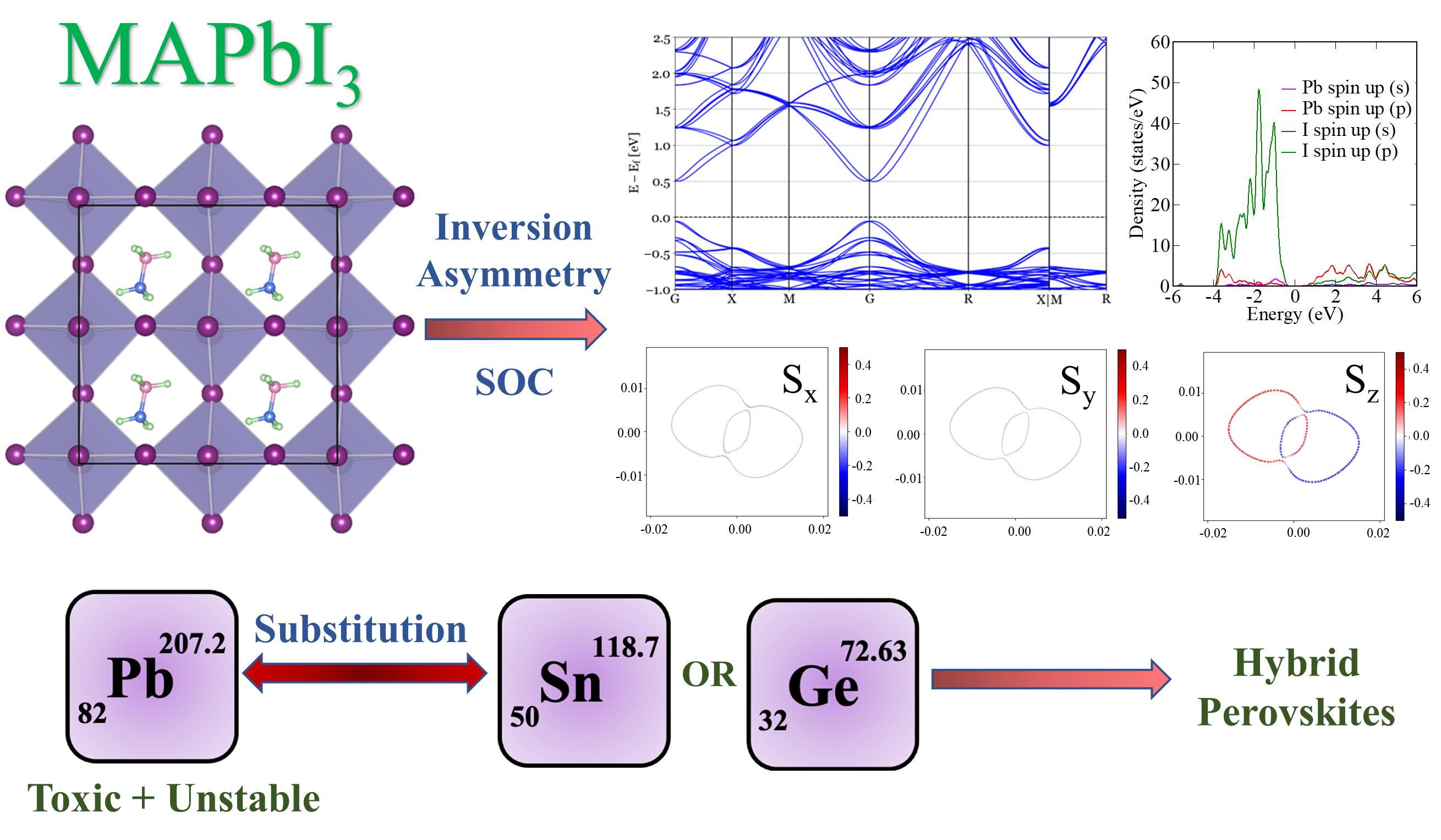}
  \end{figure}	
  \end{tocentry}
\end{abstract}

In recent years, inorganic-organic lead halide perovskite compounds, particularly methylammonium lead halide perovskite (MAPbI$_3$), have attracted a great attention from the scientific community. They have the potential of revolutionizing the field of optoelectronics and spintronics owing to their exotic properties like long diffusion length, appropriate band gaps, high carrier mobility, high absorption coefficient and cheap cost of manufacturing.\cite{snaith2013perovskites,green2014emergence,gratzel2014light,kojima2009organometal,lee2012efficient,xing2013long,manser2016intriguing,swarnkar2016quantum,stranks2015metal,nam2018methodologies,tombe2017optical,burschka2013sequential} In comparison to the first experimental work, where the power conversion efficiency (PCE) of perovskite solar cells was only 3.8$\%$\cite{kojima2009organometal}, intense efforts in past years have led to an impressive increase of PCE to over 25$\%$.\cite{nrel,sahli2018fully} In addition to this, because of the presence of heavy element Pb, there is a large spin-orbit coupling (SOC) which influences the band structure of these lead halide perovskites.\cite{even2012electronic,even2013importance} Fascinatingly, SOC in association with inversion asymmetry, induces various exotic phenomena like topological surface states\cite{schindler2018higher}, persistent spin textures\cite{tao2018persistent,zhao2020purely} and Rashba-Dresselhaus (RD) effects.\cite{sandoval2012rashba,ponet2018first,di2016intertwined,yamauchi2015coupling,tao2017reversible} In case of crystals where center of inversion is not present, there is a finite gradient of potential and hence a Lorentz transformed magnetic field acts on relativistically moving electrons. So, at non-time-reversal-invariant \emph{k}-points, the degenerate bands split into states of opposite spins, lifting Kramer’s degeneracy. The orientation of spin is decided by spin-orbit field, which depends on momentum. Dresselhaus\cite{dresselhaus1955spin} studied that for crystals without a centre of inversion i.e. bulk inversion asymmetry (BIA), SOC lifts spin degeneracy of the electronic bulk states and the effect was originally reported in zinc-blende. Rashba\cite{rashba1960properties} studied spin-orbit effects in two-dimensional crystals in an asymmetric potential referred to as structural inversion asymmetry (SIA) and the effect was initially reported in wurtzite. The differentiator between these effects is the origin of the non-centrosymmetry. RD effect has great applications in spintronics,\cite{di2013electric,plekhanov2014engineering} and thus, lead halide perovskites are seen as promising materials for applications in the field.\cite{kim2014switchable,kutes2014direct,giovanni2015highly,kepenekian2017rashba,rinaldi2018ferroelectric}

In the bulk of MAPbI$_3$, the inversion symmetry is broken by two major mechanisms: static and dynamic. In the static one, the octahedron PbI framework gets distorted due to small and long bond lengths.\cite{rakita2017tetragonal} The dynamic one is where the PbI framework is fixed but inversion asymmetry comes into the picture due to the rotation of MA ions.\cite{frohna2018inversion} Dynamic structural fluctuations can even take place because of  phonon modes or interaction of MA ions with the PbI framework.\cite{etienne2016dynamical,motta2015revealing,wu2017light} Also, the system has strong SOC due to the presence of heavy element iodine in addition to lead.\cite{frohna2018inversion} However, Pb has toxic nature that hampers MAPbI$_3$ from its practical uses. Additionally, under air and moisture exposure, Pb is intrinsically unstable and ultimately oxidizes from Pb$^2$$^+$ to Pb$^4$$^+$, reducing the performance of devices based on MAPbI$_3$.\cite{abdelmageed2018effect,sun2016mixed} Thus, it is essential to reduce the Pb extent in the compound by substituting appropriate alternatives and making a hybrid perovskite. Pb extent can also be reduced by creating Pb-vacancies in the compound, however, they  are not stable at all.\cite{basera2020reducing}

MAPbI$_3$ crystals exhibit multiple phases: cubic with a space group Pm$\bar{3}$m above 327 K\cite{kawamura2002structural}, tetragonal with two debatable space groups, i.e, centrosymmetric I4/mcm (point group 4/mmm) and non-centrosymmetric I4cm (point group 4mm) between 162 and 327 K\cite{kawamura2002structural,weller2015complete,stoumpos2013semiconducting,rakita2017tetragonal}, and orthorhombic with a space group Pnma below 162 K\cite{weller2015complete}. However, in case of solar cell applications, just the tetragonal room temperature phase and the cubic high-temperature phase are of relevance. In our present work, we studied the cubic phase of MAPbI$_3$  using state-of-the-art density functional theory (DFT)\cite{hohenberg1964inhomogeneous,kohn1965self} and we intend to reduce Pb extent by substituting Sn and Ge in the perovskite and parallelly maintain spin-splitting effects.   

Vienna ab initio Simulation Package (VASP)\cite{PhysRevB.47.558,PhysRevB.54.11169} is used to perform DFT calculations. The projector augmented wave (PAW) pseudopotential method,\cite{blochl1994projector,kresse1999ultrasoft} as implemented in VASP, is used to describe all ion-electron interactions in elemental constituents. The exchange-correlation ($\epsilon_{xc}$) functionals used for DFT calculations are Perdew-Burke-Ernzerhof (PBE)\cite{perdew1996generalized} and the non-local Heyd–Scuseria–Ernzerhof (HSE06)\cite{heyd2003hybrid} $\epsilon_{xc}$ functionals with SOC. We have increased the size of the supercell till the state of single defect is completely localized with periodic boundary conditions. The converged supercell (2×2×2) contains 96 atoms, i.e., MA$_8$Pb$_8$I$_{24}$. Since in this supercell, there are 8 Pb atoms, so all calculations are done until all the 8 Pb atoms are replaced one by one by Sn/Ge. During optimization of the structures, the total energy difference between two ionic relaxation steps is kept lesser than 0.0001 eV and the tolerance on forces between two steps is kept 0.01 eV/ Å. A 4×4×4 Monkhorst \emph{k}-mesh size is used for optimization. For energy calculations, the \emph{k}-mesh is converged at 4×4×4. The plane wave energy cutoff is kept 520 eV for all the calculations. 

Firstly, formation energies are calculated using PBE+SOC and HSE06+SOC $\epsilon_{xc}$ functionals. It is known that HSE06+SOC is more accurate for the calculations as it determines the position of energy bands more precisely, yet it is interesting to verify whether PBE works well in the context of thermodynamic stability. Then, bandstructures, band gaps, electronic atom-projected partial density of states (pDOS), spin textures and Rashba coefficients are computed using PBE+SOC $\epsilon_{xc}$ functional. PyProcar\cite{herath2020pyprocar} is used to plot band structures and calculate the constant energy contour plots of spin textures. For all these other calculations, in view of computational cost, we have used PBE+SOC because in the present work, we are more interested to know the trends on successive substitutions in comparison to the pristine structure. 

As stated above, there are two major factors breaking inversion symmetry, i.e., distortion of PbI framework and rotation of MA ions. We have quantified the contribution of each factor by doing all calculations under (i) complete relaxation of structures and (ii) selective dynamics where the PbI framework was fixed (see Figure 1 a, b). Thus, the completely relaxed structures include effects due to both factors and the structures relaxed under selective dynamics include only MA ion rotation effects. Hence, the effects due to only PbI framework distortion can be seen as a difference of both calculations.The PbI framework is non-centrosymmetric in the completely relaxed structures and centrosymmetric in the MA-relaxed structures, which are, however overall non-centrosymmetric due to the random MA orientations.

Also, we calculated change in Rashba parameters on application of axial strain using PBE+SOC $\epsilon_{xc}$ functional. This calculation was done for completely relaxed structures only, keeping in view the comparatively small contribution to Rashba coefficient obtained from only MA relaxed unstrained structures. Lastly, mechanical properties were calculated for all configurations using PBE+SOC $\epsilon_{xc}$ functional.

\begin{center}
\begin{figure}[H]
	\includegraphics[width=17cm]{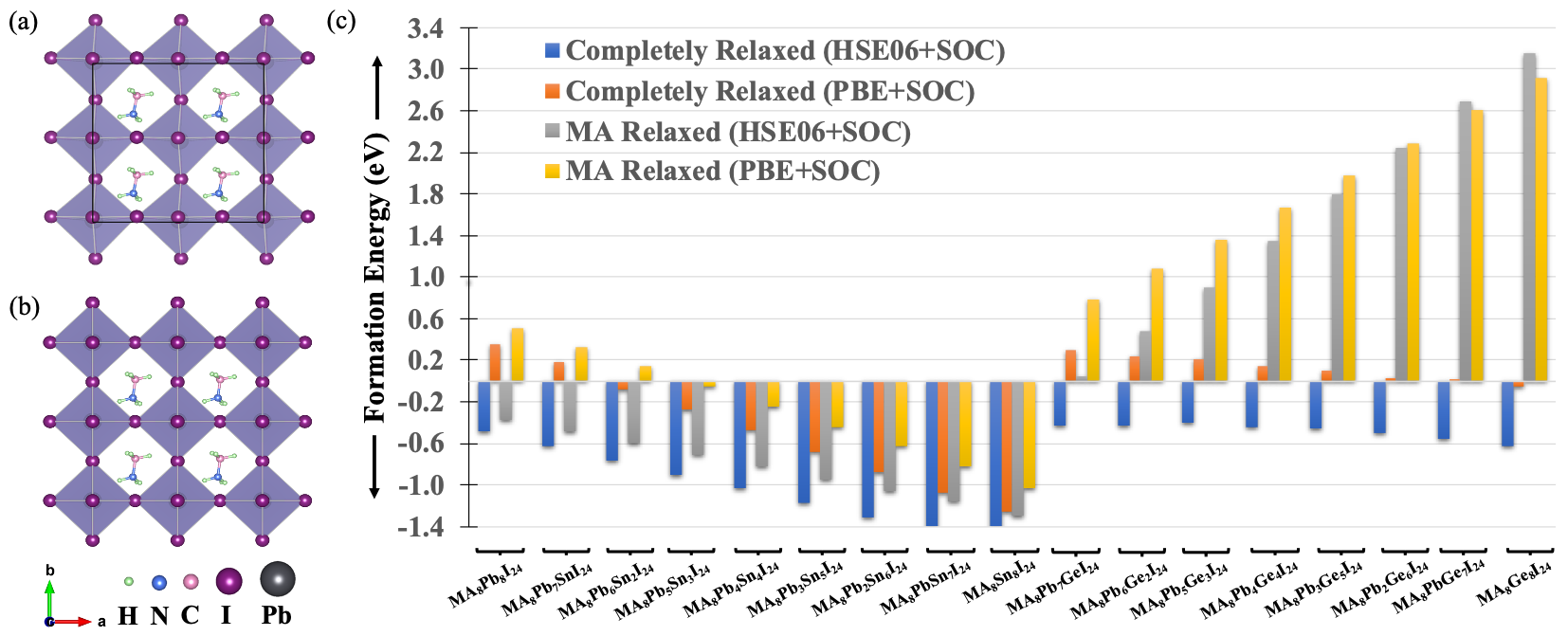}
	\caption{Optimized crystal structures of MAPbI$_3$ in the cubic phase (a) when all atoms are allowed to relax and (b) when only MA ions are allowed to relax. Structure in (a) exhibits distortions of the PbI framework as opposed to structure in (b). (c) Formation energies for completely and MA relaxed pristine and Sn/Ge substituted structures calculted using HSE06+SOC and PBE+SOC $\epsilon_{xc}$ functionals.} 
	\label{1}
\end{figure}
\end{center}

We have calculated formation energies of all structures to find out whether they are thermodynamically stable or not. To calculate formation energy \emph{E$_f (x)$} of MAPb$_{1-x}$Sn$_x$I$_3$ (in case of 2×2×2 supercell, i.e., MA$_8$Pb$_{8-x}$Sn$_x$I$_{24}$), we considered total energies of precursor materials, i.e., PbI$_2$, MAI and SnI$_2$ and subtracted them from the energy of original material according to the following equation: 
\begin{equation}
	\begin{aligned}
			E_f (x) = E\mathrm{(MA_8Pb_{8-\textit{x}}Sn_\textit{x}I_{24})} - 8 E\mathrm{(MAI)} - x E\mathrm{(SnI_2)} - (8 - x) E\mathrm{(PbI_2)}
			\label{eq1}
	\end{aligned}
\end{equation}

Similar to Sn, to calculate formation energy \emph{E$_f(x)$} of MAPb$_{1-x}$Ge$_x$I$_3$ (in case of 2×2×2 supercell, i.e., MA$_8$Pb$_{8-x}$Ge$_x$I$_{24}$), we considered total energies of precursor materials, i.e., PbI$_2$, MAI and GeI$_2$ and subtracted them from the energy of original material according to the following equation:
\begin{equation}
	\begin{aligned}
		E_f (x) = E\mathrm{(MA_8Pb_{8-\textit{x}}Ge_\textit{x}I_{24})} - 8 E\mathrm{(MAI)} - x E\mathrm{(GeI_2)} - (8 - x) E\mathrm{(PbI_2)}
		\label{eq2}
	\end{aligned}
\end{equation}

Note that, in above equations, the coefficients of all terms are taken such that the number of MA, Pb, Sn/Ge and I atoms are balanced stoichiometrically. 

Figure 1 c shows comparison of formation energies of pristine and Sn/Ge substituted hybrid structures using different $\epsilon_{xc}$ functionals and relaxation mechanisms (for exact values, see section I in Supporting Information (SI)). The pristine structure is found to be unstable using PBE+SOC, which is not the case.\cite{qiu2019room} This is due to incorrect positions of valence band maxima and conduction band minima. Therefore, we have calculated the formation energies using HSE06+SOC. The negative values obtained using HSE06+SOC for all structures indicate that they are thermodynamical stable. Besides this, the fully relaxed structures are more stable as compared to the corresponding only MA relaxed structures in each case. The contribution of PbI framework distortion towards formation energy can be analyzed as a difference of corresponding formation energies of completely relaxed and only MA relaxed structures in each case.

Then, we have plotted bandstructures inside first Brillouin zone, taking into consideration the high symmetry path. Firstly, we have performed non-spin-polarized calculations for the pristine structure. After that, we have considered SOC in the calculation of electronic bandstructures for pristine and substituted structures. All calculations were performed using both relaxation mechanisms, as mentioned before (See section II in SI for figures). The splitting of bands can be clearly observed at the high symmetry point G in all the structures after considering SOC. This removes the spin degeneracy of valence and conduction bands, leading to the formation of distinct parabolic band minima shifted from the high-symmetry point G of the brillouin zone. The amount of splitting is different in the directions M-G and G-R. The bandstructures of the fully relaxed structures exhibit a large splitting of the band edges. On contrast, in the MA-relaxed structures, the splitting is considerably reduced in all cases. Also, the comparison of bandstructures of pristine and Sn-doped structures shows that the bandgap decreases upon substitution upto five Sn atoms and then again starts to increase. However, it is always lesser in doped structures when compared with the pristine structure. On Sn substitution, because of strong s-p and p-p couplings, there is a downward shift in the lower part of conduction band. This indicates that the optical absorption coefficient is high and is a good indicator for solar cell application. On the other hand, a comparison of bandstructures of pristine and doped structures shows that the bandgap increases in Ge-substituted structures. The values of band gaps are listed in Table 1.

\begin{table}[htbp]
	\caption {Band gaps (in eV) of MA$_8$Pb$_{8-x}$\{Sn/Ge\}$_x$I$_{24}$ perovskites for completely and MA relaxed structures. For pristine structure, band gap is calculated with and without SOC, whereas, for doped structures, SOC is taken into consideration.}
	\begin{center}
		\begin{adjustbox}{width=0.6\textwidth} 
			\setlength\extrarowheight{+4pt}
			\begin{tabular}[c]{|c|c|c|} \hline		
				\textbf{Configurations} & \textbf{Completely Relaxed} & \textbf{MA Relaxed} \\ \hline
				MA$_8$Pb$_8$I$_{24}$ (without SOC)         &  	1.59      &  	1.45  
				          \\ \hline
					MA$_8$Pb$_8$I$_{24}$ (with SOC)     &  	   0.55
				&  	    0.35
				\\ \hline          
				MA$_8$Pb$_7$SnI$_{24}$     & 	0.49     &  	 0.30
				    \\ \hline
				MA$_8$Pb$_6$Sn$_2$I$_{24}$     & 0.48     &  0.29	        \\ \hline
				MA$_8$Pb$_5$Sn$_3$I$_{24}$     & 0.42   &  	0.27        \\ \hline
				MA$_8$Pb$_4$Sn$_4$I$_{24}$    & 0.38    &  	 0.24       \\ \hline
				MA$_8$Pb$_3$Sn$_5$I$_{24}$     & 0.35    &  	 0.22        \\ \hline
				MA$_8$Pb$_2$Sn$_6$I$_{24}$     & 0.36     &  0.23	         \\ \hline
				MA$_8$PbSn$_7$I$_{24}$     &  0.43     &  	0.27         \\ \hline
				MA$_8$Sn$_8$I$_{24}$     & 0.47     &  	0.29           \\ \hline
				MA$_8$Pb$_7$GeI$_{24}$     &  0.69    &  	0.39      \\ \hline
				MA$_8$Pb$_6$Ge$_2$I$_{24}$     & 0.71     &  	0.42         \\ \hline
				MA$_8$Pb$_5$Ge$_3$I$_{24}$     &  0.85     &  	0.50        \\ \hline
				MA$_8$Pb$_4$Ge$_4$I$_{24}$     & 0.90     &  0.53	     \\ \hline
				MA$_8$Pb$_3$Ge$_5$I$_{24}$      &  1.00    &  0.59	        \\ \hline
				MA$_8$Pb$_2$Ge$_6$I$_{24}$    &  1.07     &  	0.63        \\ \hline
				MA$_8$PbGe$_7$I$_{24}$     &  1.19     &  	0.70        \\ \hline
				MA$_8$Ge$_8$I$_{24}$     & 1.37    &  	0.78        \\ \hline
			\end{tabular}
		\end{adjustbox}
		\label{T2}
	\end{center}
\end{table}

After that, we have calculated the projected density of states (pDOS). Firstly, we have performed non-spin-polarized calculations of the pristine structure for both relaxation mechanisms. After that, we have considered SOC in the calculation of pDOS for pristine and substituted structures for both relaxation mechanisms (See section III in SI for figures). For all the structures, p-orbital of I contributes majorly to DOS, in addition to s-orbital of I which also determines DOS to some extent. The contribution from s and p orbitals of Pb decreases and the contribution from s and p orbitals of Sn/Ge increases as number of substituted atoms increase.

The nature of band splitting due to Rashba and Dresselhaus effects is similar. However, by projection of spin-orientation in Fourier space, called as spin texture, we can know about the type of splitting. We have plotted spin textures by calculating the expectation values of spin operators S$_i$ (\emph{i = x, y, z}), which is given as, 
\begin{equation}
	\begin{aligned}
		\langle \mathrm{S}_i \rangle = \langle \Psi_k \mid \sigma_i \mid \Psi_k \rangle
		\label{eq3}
	\end{aligned}
\end{equation}
where $\sigma_i$ are the Pauli matrices, and $\Psi_k$ is the spinor eigenfunction obtained from noncollinear spin calculations. 

\begin{center}
\begin{figure}[H]
	\includegraphics[width=17cm]{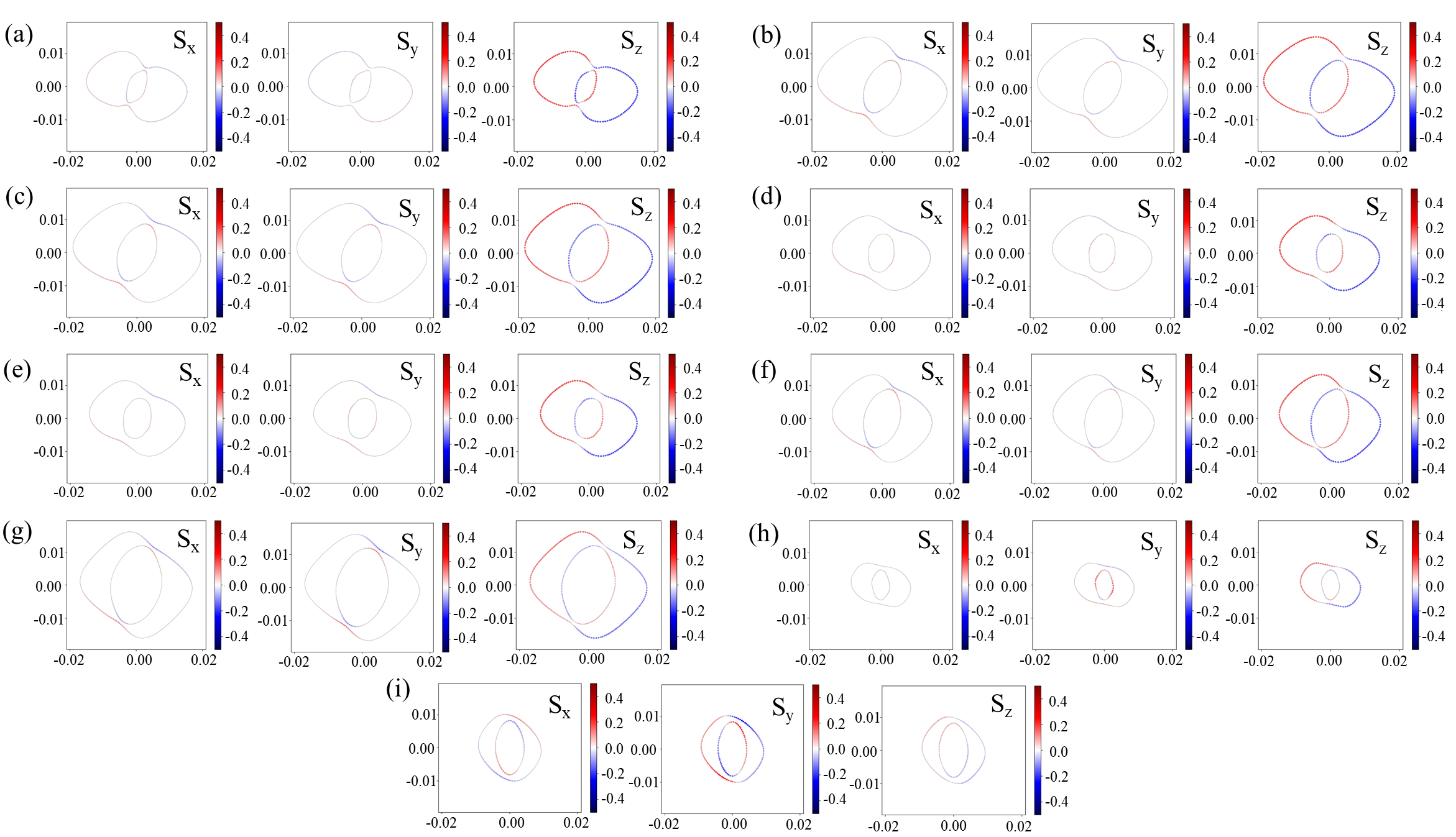}
	\caption{(a) Spin textures for completely relaxed configurations (a) MA$_8$Pb$_8$I$_{24}$ (b) MA$_8$Pb$_7$SnI$_{24}$ (c) MA$_8$Pb$_6$Sn$_2$I$_{24}$ (d) MA$_8$Pb$_5$Sn$_3$I$_{24}$ (e) MA$_8$Pb$_4$Sn$_4$I$_{24}$ (f) MA$_8$Pb$_3$Sn$_5$I$_{24}$ (g) MA$_8$Pb$_2$Sn$_6$I$_{24}$ (h) MA$_8$PbSn$_7$I$_{24}$ (i) MA$_8$Sn$_8$I$_{24}$.}
	\label{1}
\end{figure}
\end{center}

\begin{center}
\begin{figure}[H]
	\includegraphics[width=17cm]{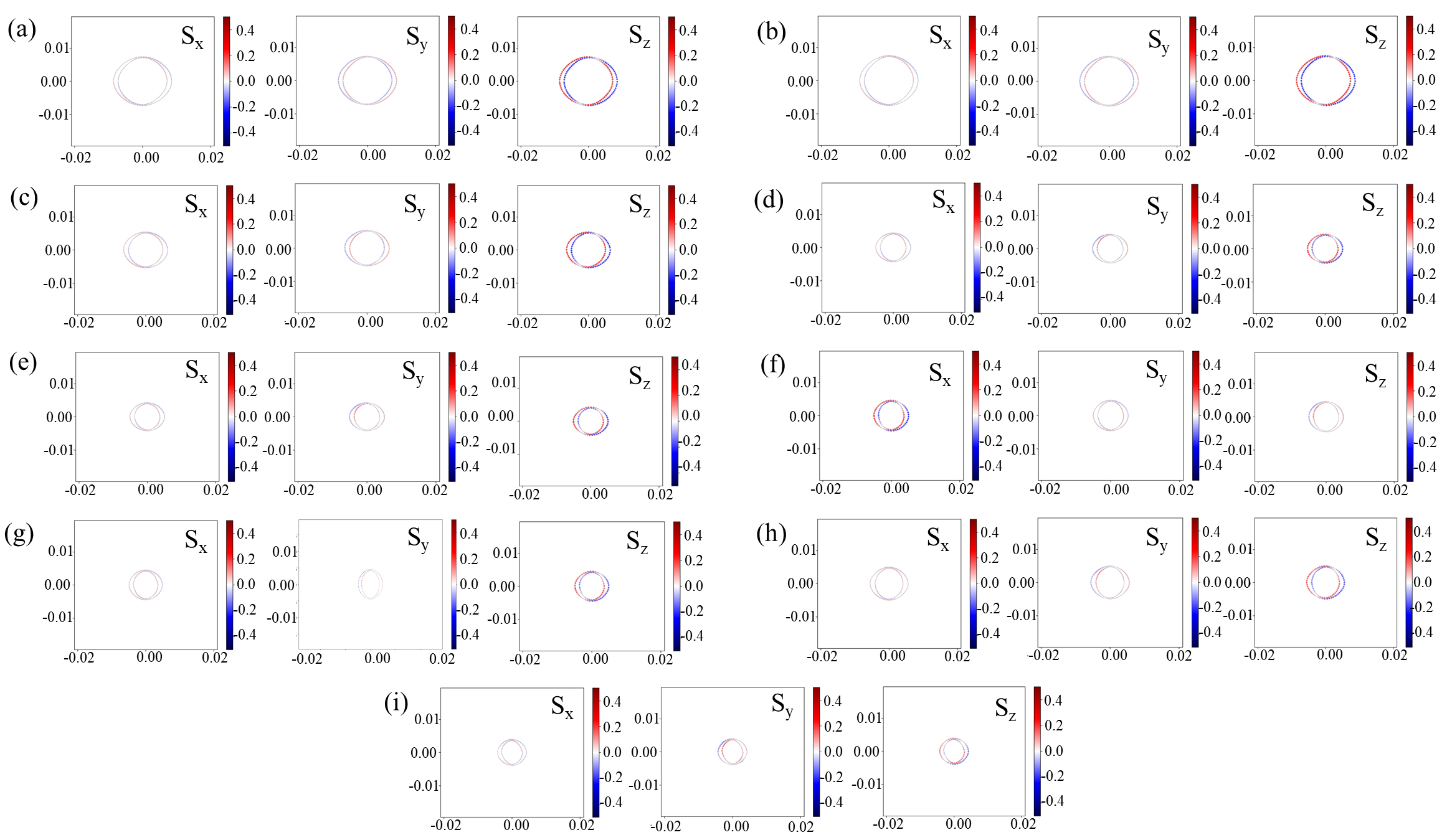}
	\caption{(a) Spin textures for MA relaxed configurations (a) MA$_8$Pb$_8$I$_{24}$ (b) MA$_8$Pb$_7$SnI$_{24}$ (c) MA$_8$Pb$_6$Sn$_2$I$_{24}$ (d) MA$_8$Pb$_5$Sn$_3$I$_{24}$ (e) MA$_8$Pb$_4$Sn$_4$I$_{24}$ (f) MA$_8$Pb$_3$Sn$_5$I$_{24}$ (g) MA$_8$Pb$_2$Sn$_6$I$_{24}$ (h) MA$_8$PbSn$_7$I$_{24}$ (i) MA$_8$Sn$_8$I$_{24}$.}
	\label{1}
\end{figure}
\end{center}

Figure 2, 3, 4, 5 depict constant energy 2D surface spin textures for completely relaxed and MA relaxed pristine and Sn/Ge hybrid perovskites. The spin textures for pristine and doped structures have similar nature for a lesser number of Sn/Ge atoms substituted in the supercell. Here, only the spin  component S$_z$ is dominant in these structures. This indicates that the major splitting is Rashba splitting for lower substitutions. However, for higher substitutions, Dresselhaus effect also comes into picture as evident from spin textures. The impact of PbI framework distortion can be clearly seen on the nature of spin textures when we compare fully relaxed and MA relaxed structures.
\begin{center}
\begin{figure}[H]
	\includegraphics[width=17cm]{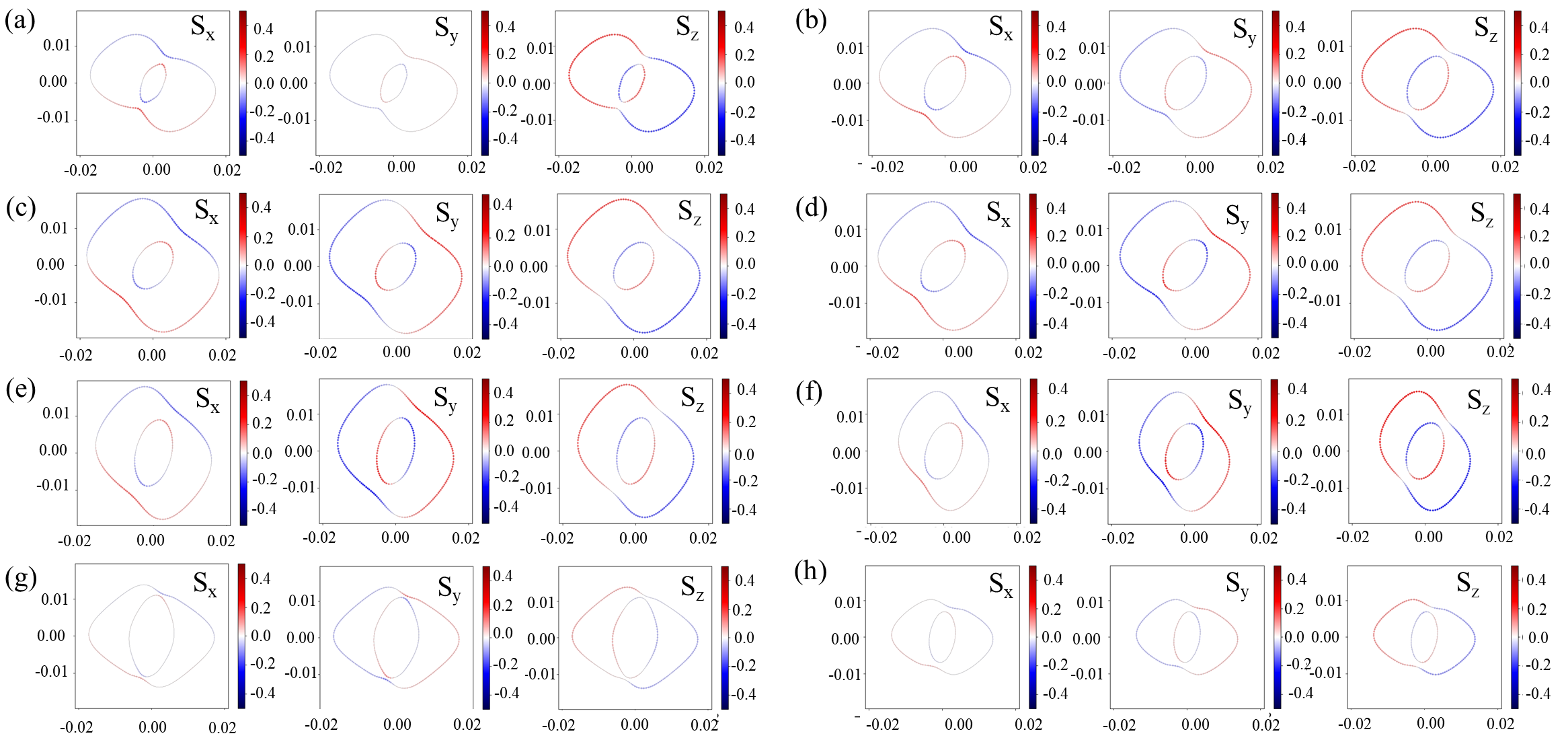}
	\caption{(a) Spin textures for completely relaxed configurations (a) MA$_8$Pb$_7$GeI$_{24}$ (b) MA$_8$Pb$_6$Ge$_2$I$_{24}$ (c) MA$_8$Pb$_5$Ge$_3$I$_{24}$ (d) MA$_8$Pb$_4$Ge$_4$I$_{24}$ (e) MA$_8$Pb$_3$Ge$_5$I$_{24}$ (f) MA$_8$Pb$_2$Ge$_6$I$_{24}$ (g) MA$_8$PbGe$_7$I$_{24}$ (h) MA$_8$Ge$_8$I$_{24}$.}
	\label{1}
\end{figure}
\end{center}

\begin{center}
\begin{figure}[H]
	\includegraphics[width=17cm]{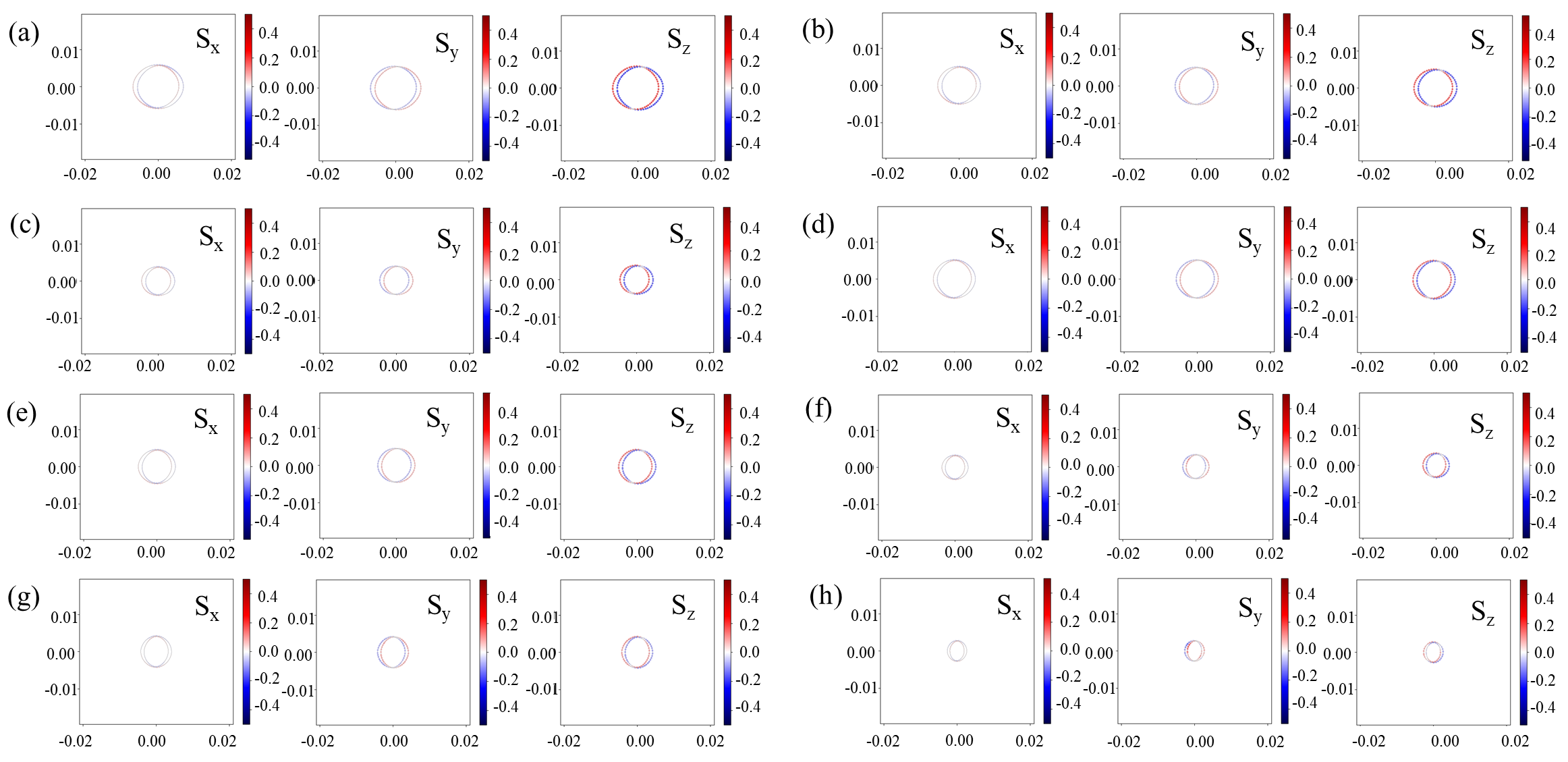}
	\caption{(a) Spin textures for MA relaxed configurations (a) MA$_8$Pb$_7$GeI$_{24}$ (b) MA$_8$Pb$_6$Ge$_2$I$_{24}$ (c) MA$_8$Pb$_5$Ge$_3$I$_{24}$ (d) MA$_8$Pb$_4$Ge$_4$I$_{24}$ (e) MA$_8$Pb$_3$Ge$_5$I$_{24}$ (f) MA$_8$Pb$_2$Ge$_6$I$_{24}$ (g) MA$_8$PbGe$_7$I$_{24}$ (h) MA$_8$Ge$_8$I$_{24}$.}
	\label{1}
\end{figure}
\end{center}
	
The values of the Rashba coefficient ($\alpha$) are calculated using the formula 2\emph{E$_R$/k$_0$}, where \emph{E$_R$} is Rashba spin-splitting energy and \emph{k$_0$} is offset momentum as shown in the schematic representation in figure 6 a, b. Figure 6 c depicts the value of Rashba coefficients for lower conduction bands in completely relaxed and MA relaxed structures, respectively, for pristine and Sn/Ge substituted structures. There are two contradictory factors which decide value of Rashba coefficient: structural inversion asymmetry and effect of less/more heavy elements present. Therefore, the value depends on the net effect of both. The Rashba coefficient calculated in doped structures in both directions (X to G and G to M) has a significant value for completely relaxed structures. Our MA-relaxed structures exhibit a small value of  Rashba Coefficient. Hence PbI framework distortion plays a significant role in spin-splitting effects (for exact values, see section IV in SI).
\begin{center}
\begin{figure}[H]
	\includegraphics[width=16cm]{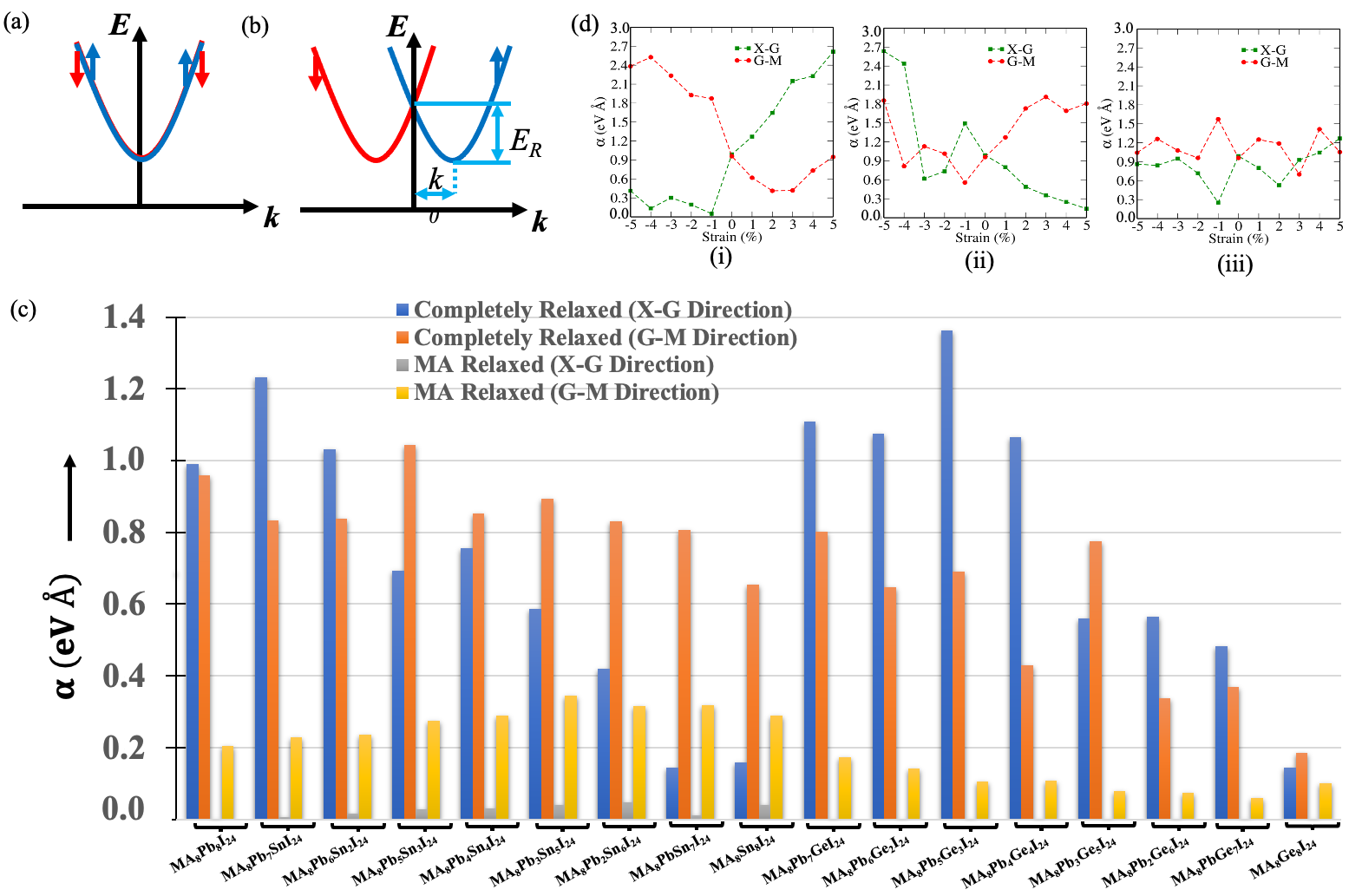}
	\caption{(a) Electron dispersion relation of a normal conduction band depicting a double spin-degenarate band with one minima. (b) Electron dispersion relation of band under subjection to Rashba splitting. (c) Rashba coefficients for lower conduction bands in completely and MA relaxed pristine and Sn/Ge substituted structures calculated in X-G and G-M directions. (d) Rashba coefficients for lower conduction bands when strain is applied along (i) ‘\emph{a}’, (ii) '\emph{b}' and (iii) '\emph{c}' axis respectively in MA$_8$Pb$_8$I$_{24}$.}
	\label{1}
\end{figure}
\end{center}

To examine the effect of axial strain on these perovskites, we have applied strain on completely relaxed structures. There are three axes along which strain can be applied in cubic phase. The electric field component in unstrained configuration is different along different axis for each configuration. We have applied strain along ‘\emph{a}’, ‘\emph{b}’ and ‘\emph{c}’ axis for pristine structure and only along ‘\emph{c}’ axis for all the substituted ones. The lattice vector \emph{a/b/c} is changed from -$5\%$ to +$5\%$ . Here “+” indicates tensile strain and and “-” indicates compressive strain. Subsequently, we have calculated Rashba coefficients for each case. We observe variation in Rashba parameters on the application of strain, indicating that the material is tunable for applications in spintronics.  Figure 6 d depicts this tunability in pristine structure when strain is applied along (a) ‘\emph{a}’, (b) '\emph{b}' and (c) '\emph{c}' axis respectively. Refer Section V in SI for details on similar observed trends in substituted structures.

To study the mechanical properties, the elastic constants of the structures were calculated. We have three independent elastic constants \textit{C$_\textit{{11}}$}, \textit{C$_\textit{{12}}$} and \textit{C$_\textit{{44}}$} for the studied structures. By using the values of these elastic constants, we calculated properties like bulk moduli (\textit{B}), shear moduli (\textit{G}), Pugh’s ratio (\textit{B/G}) and Poisson’s ratio ($\nu$). We used the Born's stability criteria for cubic phase to evaluate the mechanical stability. Then, in order to distinguish ductile materials from brittle ones, we used the two famous parameters, i.e. Pugh's ratio and Poisson's ratio. We found that the elastic constants for MA$_8$Pb$_8$I$_{24}$, MA$_8$Pb$_6$Sn$_2$I$_{24}$, MA$_8$Pb$_5$Sn$_3$I$_{24}$, MA$_8$Pb$_4$Sn$_4$I$_{24}$, MA$_8$Pb$_2$Sn$_6$I$_{24}$, MA$_8$Pb$_6$Ge$_2$I$_{24}$, MA$_8$Pb$_5$Ge$_3$I$_{24}$ and MA$_8$Pb$_2$Ge$_6$I$_{24}$ satisfy the Born's stability criteria, implying that these materials are mechanically stable. Further, among these materials, it was found that only MA$_8$Pb$_8$I$_{24}$,  MA$_8$Pb$_4$Sn$_4$I$_{24}$,  MA$_8$Pb$_2$Sn$_6$I$_{24}$, MA$_8$Pb$_4$Ge$_2$I$_{24}$ and MA$_8$Pb$_2$Ge$_3$I$_{24}$ are ductile.
The equations used and the data for mechanically stable materials is provided in 
Section VI of SI.

In conclusion, we have performed relativistic first-principles density functional theory calculations to study Rashba and Dresselhaus effects in cubic phase of MAPbI$_3$ with Pm$\bar{3}$m space group symmetry when it is substituted with tin and germanium. All the structures when substituted with different number of tin and germanium atoms in our supercell were found to be thermodynamically stable. Due to  significant amount of SOC, spin-splitting effects are observed around the \emph{k}-point G. In Sn-doped structures, the band gap was found to be lesser when compared to pristine structure which is an indicator of high absorption coefficient. On the other hand, in Ge doped structures, bandgap was found to be more when compared to pristine structure. The spin textures indicated the Rashba-type of spin-splitting for lesser number of Sn or Ge atoms substituted in pristine supercell. Also, it was observed that in addition to Rashba effect, Dresselhaus effect comes into picture for higher substitutions. A significant value of Rashba coefficient was observed in substituted structures. Also strain was applied and values of Rashba coefficients were found which indicated that our devices can be tuned for various applications. Moreover, some of the doped systems are found to become ductile materials. MAPbI$_3$ is already a promising material for spintronics, optoelectronics and spin-orbitronics because of its unique combination of useful properties which are difficult to be found even in the best semiconductors of single crystals, inspite of being made from solution processing methods. Now, replacing lead with tin or germanium in devices based on MAPbI$_3$ can prove to be a game changer. It reduces toxicity and unstability (under air and moisture) encountered due to lead and parallely gives us the desired RD effects. 

\begin{acknowledgement}
	M.A. acknowledges High Performance Computing (HPC) facility  at IIT Delhi for computational resources. P.B. acknowledges UGC, India, for the senior research fellowship [grant no. 1392/(CSIR-UGC NET JUNE 2018)]. S.S. acknowledges CSIR, India, for the senior research fellowship [grant no. 09/086(1432)/2019-EMR-I]. Suraj acknowledges HPC facility  at IIT Delhi for computational resources. S.B. acknowledges financial support from SERB under a core research grant (Grant No. CRG/2019/000647) to set up his HPC facility “Veena” at IIT Delhi for computational resources.
\end{acknowledgement}

\begin{suppinfo}
	See supplementary material for the details of formation energies, bandstructures, band gaps, pDOS (projected density of states), mechanical properties, Rashba parameters and their variation on application of axial strain on MA$_8$Pb$_{8-x}$\{Sn/Ge\}$_x$I$_{24}$ perovskites.
\end{suppinfo}
	
\providecommand{\latin}[1]{#1}
\makeatletter
\providecommand{\doi}
{\begingroup\let\do\@makeother\dospecials
	\catcode`\{=1 \catcode`\}=2 \doi@aux}
\providecommand{\doi@aux}[1]{\endgroup\texttt{#1}}
\makeatother
\providecommand*\mcitethebibliography{\thebibliography}
\csname @ifundefined\endcsname{endmcitethebibliography}
{\let\endmcitethebibliography\endthebibliography}{}

\end{document}


\begin{center}
{\Large \bf Supplemental Material}\\ 
\end{center}
\begin{enumerate}[\bf I.]
	
\item Formation energies of MA$_8$Pb$_{8-x}$\{Sn/Ge\}$_x$I$_{24}$ perovskites using HSE06+SOC and PBE+SOC exchange-correlation ($\epsilon_{xc}$) functionals
\item Band structures of MA$_8$Pb$_{8-x}$\{Sn/Ge\}$_x$I$_{24}$ perovskites
\item Projected density of states (pDOS) of MA$_8$Pb$_{8-x}$\{Sn/Ge\}$_x$I$_{24}$ perovskites
\item Rashba coefficients of MA$_8$Pb$_{8-x}$\{Sn/Ge\}$_x$I$_{24}$ perovskites
\item Variation of Rashba parameters of MA$_8$Pb$_{8-x}$\{Sn/Ge\}$_x$I$_{24}$ perovskites on application of axial strain
\item Mechanical properties of MA$_8$Pb$_{8-x}$\{Sn/Ge\}$_x$I$_{24}$ perovskites

\end{enumerate}
\vspace*{12pt}
\clearpage

\section{ Formation energies of MA$_8$Pb$_{8-x}$\{Sn/Ge\}$_x$I$_{24}$ perovskites using HSE06+SOC and PBE+SOC $\epsilon_{xc}$ functionals}
\begin{table}[H]
	\caption {Formation energies of completely relaxed structures using HSE06+SOC $\epsilon_{xc}$ functional}
	\begin{center}
		\begin{adjustbox}{width=0.6\textwidth} 
			\setlength\extrarowheight{+4pt}
			\begin{tabular}[c]{|c|c|}  \hline		
				\textbf{Configurations} & \textbf{Formation Energies (eV)}   \\ \hline	
				MA$_8$Pb$_8$I$_{24}$      &  	-0.48      \\ \hline
				MA$_8$Pb$_7$SnI$_{24}$     & 	-0.62     \\ \hline
				MA$_8$Pb$_6$Sn$_2$I$_{24}$     & -0.76       \\ \hline
				MA$_8$Pb$_5$Sn$_3$I$_{24}$     & -0.91        \\ \hline
				MA$_8$Pb$_4$Sn$_4$I$_{24}$    &  -1.04        \\ \hline
				MA$_8$Pb$_3$Sn$_5$I$_{24}$     & -1.18        \\ \hline
				MA$_8$Pb$_2$Sn$_6$I$_{24}$     & -1.31        \\ \hline
				MA$_8$PbSn$_7$I$_{24}$     &  -1.44        \\ \hline
				MA$_8$Sn$_8$I$_{24}$     &  -1.56          \\ \hline
				MA$_8$Pb$_7$GeI$_{24}$     &  -0.44       \\ \hline
			    MA$_8$Pb$_6$Ge$_2$I$_{24}$     &  -0.43        \\ \hline
				MA$_8$Pb$_5$Ge$_3$I$_{24}$     &  -0.40       \\ \hline
				MA$_8$Pb$_4$Ge$_4$I$_{24}$     & -0.44      \\ \hline
				MA$_8$Pb$_3$Ge$_5$I$_{24}$      &  -0.46       \\ \hline
				MA$_8$Pb$_2$Ge$_6$I$_{24}$    &  -0.50       \\ \hline
				MA$_8$PbGe$_7$I$_{24}$     &  -0.55       \\ \hline
			    MA$_8$Ge$_8$I$_{24}$     &  -0.62        \\ \hline

			\end{tabular}
		\end{adjustbox}
		\label{T3}
	\end{center}
\end{table}
\begin{table}[H]
	\caption {Formation energies of completely relaxed structures using PBE+SOC $\epsilon_{xc}$ functional}
	\begin{center}
		\begin{adjustbox}{width=0.6\textwidth} 
			\setlength\extrarowheight{+4pt}
			\begin{tabular}[c]{|c|c|}  \hline		
				\textbf{Configurations} & \textbf{Formation Energies (eV)}   \\ \hline	
				MA$_8$Pb$_8$I$_{24}$      &  	0.35      \\ \hline
				MA$_8$Pb$_7$SnI$_{24}$     & 	0.18     \\ \hline
				MA$_8$Pb$_6$Sn$_2$I$_{24}$     & -0.08        \\ \hline
				MA$_8$Pb$_5$Sn$_3$I$_{24}$     & -0.28           \\ \hline
				MA$_8$Pb$_4$Sn$_4$I$_{24}$    &  -0.48       \\ \hline
				MA$_8$Pb$_3$Sn$_5$I$_{24}$     & -0.68         \\ \hline
				MA$_8$Pb$_2$Sn$_6$I$_{24}$     & -0.89            \\ \hline
				MA$_8$PbSn$_7$I$_{24}$     &  -1.08         \\ \hline
				MA$_8$Sn$_8$I$_{24}$     &  -1.26           \\ \hline
				MA$_8$Pb$_7$GeI$_{24}$     &  0.30         \\ \hline
				MA$_8$Pb$_6$Ge$_2$I$_{24}$     &  0.25        \\ \hline
				MA$_8$Pb$_5$Ge$_3$I$_{24}$     & 0.21         \\ \hline
				MA$_8$Pb$_4$Ge$_4$I$_{24}$     & 0.14      \\ \hline
				MA$_8$Pb$_3$Ge$_5$I$_{24}$      &  0.10     \\ \hline
				MA$_8$Pb$_2$Ge$_6$I$_{24}$    &  0.03        \\ \hline
				MA$_8$PbGe$_7$I$_{24}$     & 0.01      \\ \hline
				MA$_8$Ge$_8$I$_{24}$     & -0.05        \\ \hline

			\end{tabular}
		\end{adjustbox}
		\label{T3}
	\end{center}
\end{table}

\begin{table}[H]
	\caption {Formation energies of MA relaxed structures using HSE06+SOC $\epsilon_{xc}$ functional}
	\begin{center}
		\begin{adjustbox}{width=0.6\textwidth} 
			\setlength\extrarowheight{+4pt}
			\begin{tabular}[c]{|c|c|}  \hline		
				\textbf{Configurations} & \textbf{Formation Energies (eV)}   \\ \hline	
				MA$_8$Pb$_8$I$_{24}$      &  	-0.37      \\ \hline
				MA$_8$Pb$_7$SnI$_{24}$     & 	-0.48     \\ \hline
				MA$_8$Pb$_6$Sn$_2$I$_{24}$     & -0.59        \\ \hline
				MA$_8$Pb$_5$Sn$_3$I$_{24}$     & -0.71        \\ \hline
				MA$_8$Pb$_4$Sn$_4$I$_{24}$    &  -0.82        \\ \hline
				MA$_8$Pb$_3$Sn$_5$I$_{24}$     & -0.94         \\ \hline
				MA$_8$Pb$_2$Sn$_6$I$_{24}$     & -1.06         \\ \hline
				MA$_8$PbSn$_7$I$_{24}$     &  -1.16         \\ \hline
				MA$_8$Sn$_8$I$_{24}$     & -1.29           \\ \hline
				MA$_8$Pb$_7$GeI$_{24}$     &  0.05      \\ \hline
				MA$_8$Pb$_6$Ge$_2$I$_{24}$     & 0.48         \\ \hline
				MA$_8$Pb$_5$Ge$_3$I$_{24}$     &  0.90        \\ \hline
				MA$_8$Pb$_4$Ge$_4$I$_{24}$     & 1.35     \\ \hline
				MA$_8$Pb$_3$Ge$_5$I$_{24}$      &  1.80        \\ \hline
				MA$_8$Pb$_2$Ge$_6$I$_{24}$    &  2.25        \\ \hline
				MA$_8$PbGe$_7$I$_{24}$     &  2.69        \\ \hline
				MA$_8$Ge$_8$I$_{24}$     &  3.16        \\ \hline

			\end{tabular}
		\end{adjustbox}
		\label{T3}
	\end{center}
\end{table}

\begin{table}[H]
	\caption {Formation energies of MA relaxed structures using PBE+SOC$\epsilon_{xc}$ functional}
	\begin{center}
		\begin{adjustbox}{width=0.6\textwidth} 
			\setlength\extrarowheight{+4pt}
			\begin{tabular}[c]{|c|c|}  \hline		
				\textbf{Configurations} & \textbf{Formation Energies (eV)}   \\ \hline	
				MA$_8$Pb$_8$I$_{24}$      &  	0.50      \\ \hline
				MA$_8$Pb$_7$SnI$_{24}$     & 	0.32       \\ \hline
				MA$_8$Pb$_6$Sn$_2$I$_{24}$     & 0.14         \\ \hline
				MA$_8$Pb$_5$Sn$_3$I$_{24}$     & -0.05            \\ \hline
				MA$_8$Pb$_4$Sn$_4$I$_{24}$    &  -0.24        \\ \hline
				MA$_8$Pb$_3$Sn$_5$I$_{24}$     & -0.44       \\ \hline
				MA$_8$Pb$_2$Sn$_6$I$_{24}$     & -0.63             \\ \hline
				MA$_8$PbSn$_7$I$_{24}$     &  -0.83         \\ \hline
				MA$_8$Sn$_8$I$_{24}$     &  -1.03          \\ \hline
				MA$_8$Pb$_7$GeI$_{24}$     &  0.79         \\ \hline
				MA$_8$Pb$_6$Ge$_2$I$_{24}$     & 1.08         \\ \hline
				MA$_8$Pb$_5$Ge$_3$I$_{24}$     & 1.37         \\ \hline
				MA$_8$Pb$_4$Ge$_4$I$_{24}$     & 1.67       \\ \hline
				MA$_8$Pb$_3$Ge$_5$I$_{24}$      &  1.98      \\ \hline
				MA$_8$Pb$_2$Ge$_6$I$_{24}$    &  2.29        \\ \hline
				MA$_8$PbGe$_7$I$_{24}$     & 2.60      \\ \hline
				MA$_8$Ge$_8$I$_{24}$     & 2.92         \\ \hline

			\end{tabular}
		\end{adjustbox}
		\label{T3}
	\end{center}
\end{table}

\newpage

\section{Band structures of MA$_8$Pb$_{8-x}$\{Sn/Ge\}$_x$I$_{24}$ perovskites}
\begin{figure}[H]
\centering
	\includegraphics[width=8cm, height=8cm]{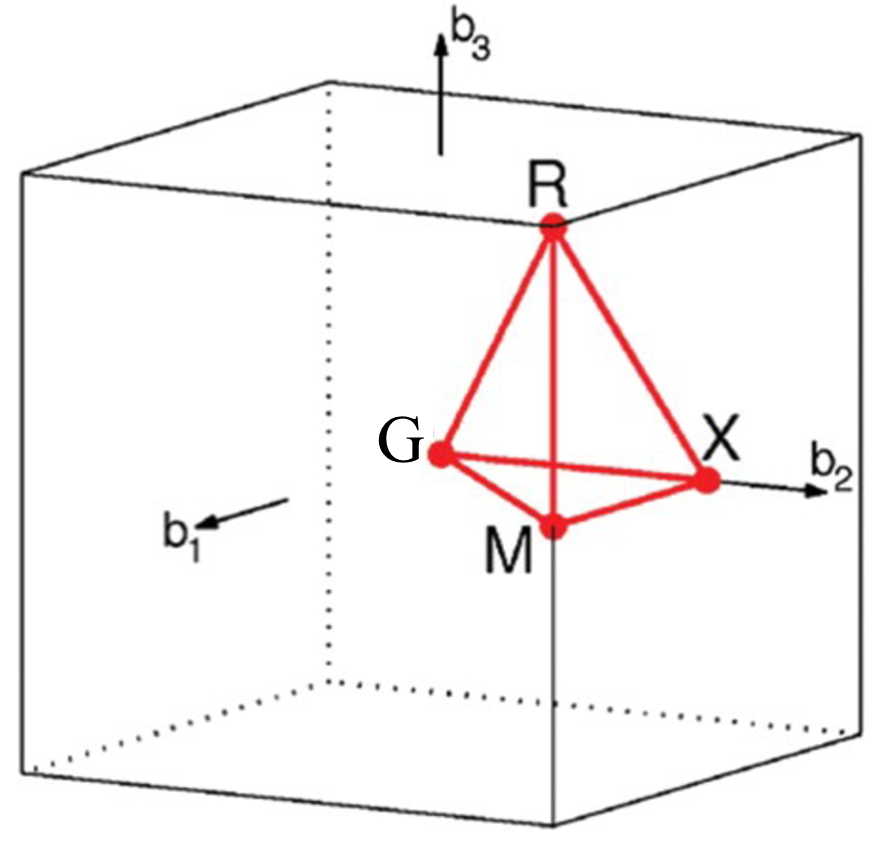}
	\caption{The first cubic brillouin zone showing the high symmetry path for band structure calculations.}
	\label{1}
\end{figure}
\begin{figure}[H]
\centering
	\includegraphics[width=15cm, height=20cm]{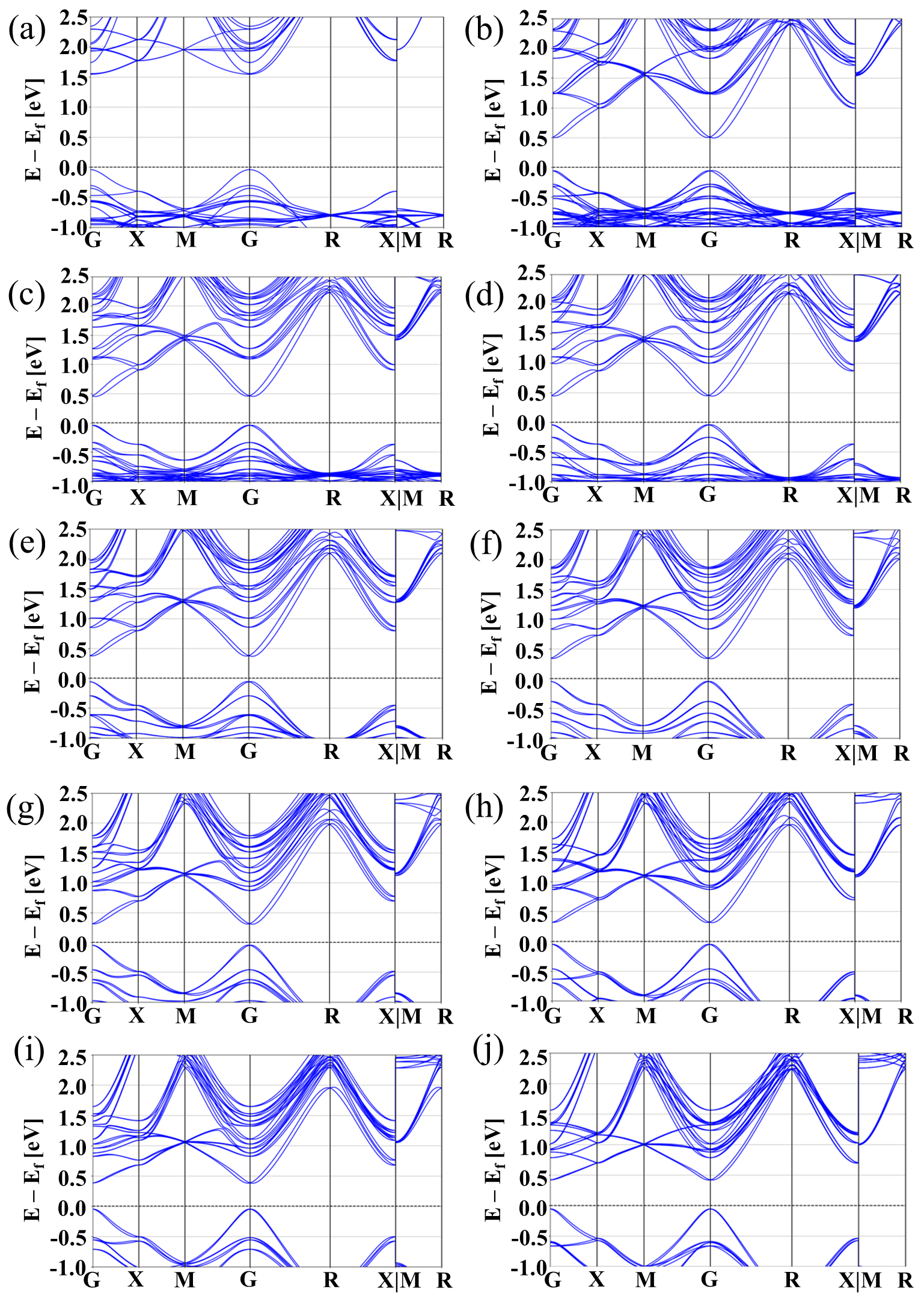}
	\caption{Electronic band structures for completely relaxed configurations calculated using PBE $\epsilon_{xc}$ functional for (a) MA$_8$Pb$_8$I$_{24}$ and using PBE+SOC $\epsilon_{xc}$ functional for (b) MA$_8$Pb$_8$I$_{24}$ (c) MA$_8$Pb$_7$SnI$_{24}$ (d) MA$_8$Pb$_6$Sn$_2$I$_{24}$ (e) MA$_8$Pb$_5$Sn$_3$I$_{24}$ (f) MA$_8$Pb$_4$Sn$_4$I$_{24}$ (g) MA$_8$Pb$_3$Sn$_5$I$_{24}$ (h) MA$_8$Pb$_2$Sn$_6$I$_{24}$ (i) MA$_8$PbSn$_7$I$_{24}$ (j) MA$_8$Sn$_8$I$_{24}$.}
	\label{1}
\end{figure}
\begin{figure}[H]
\centering
	\includegraphics[width=15cm, height=20cm]{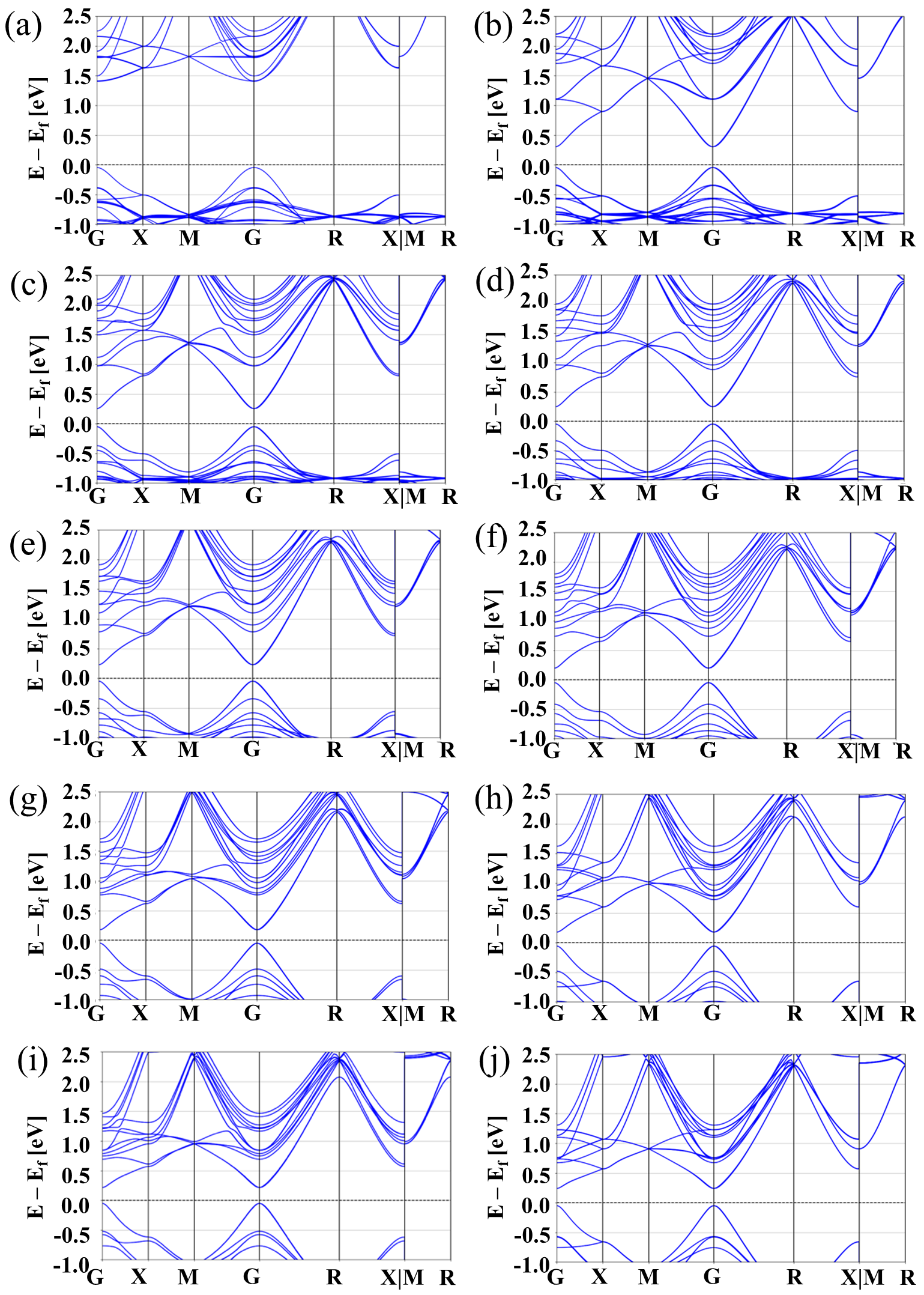}
	\caption{Electronic band structures for MA relaxed configurations calculated using PBE $\epsilon_{xc}$ functional for (a) MA$_8$Pb$_8$I$_{24}$ and using PBE+SOC $\epsilon_{xc}$ functional for (b) MA$_8$Pb$_8$I$_{24}$ (c) MA$_8$Pb$_7$SnI$_{24}$ (d) MA$_8$Pb$_6$Sn$_2$I$_{24}$ (e) MA$_8$Pb$_5$Sn$_3$I$_{24}$ (f) MA$_8$Pb$_4$Sn$_4$I$_{24}$ (g) MA$_8$Pb$_3$Sn$_5$I$_{24}$ (h) MA$_8$Pb$_2$Sn$_6$I$_{24}$ (i) MA$_8$PbSn$_7$I$_{24}$ (j) MA$_8$Sn$_8$I$_{24}$.}
	\label{1}
\end{figure}
\begin{figure}[H]
\centering
	\includegraphics[width=15cm, height=20cm]{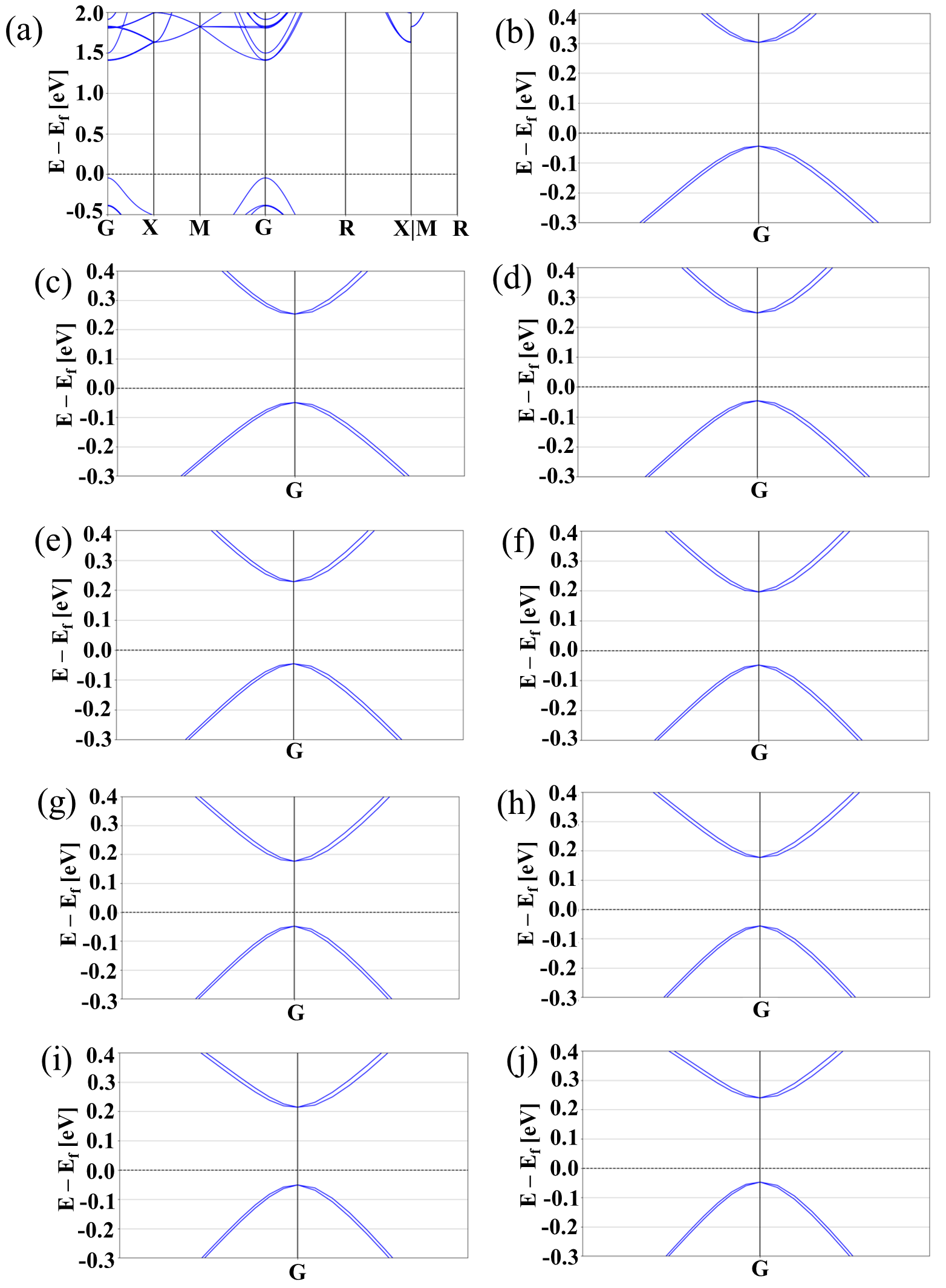}
	\caption{Enlarged view of electronic band structures for MA relaxed configurations around the high-symmetry point G calculated using PBE $\epsilon_{xc}$ functional for (a) MA$_8$Pb$_8$I$_{24}$ and using PBE+SOC $\epsilon_{xc}$ functional for (b) MA$_8$Pb$_8$I$_{24}$ (c) MA$_8$Pb$_7$SnI$_{24}$ (d) MA$_8$Pb$_6$Sn$_2$I$_{24}$ (e) MA$_8$Pb$_5$Sn$_3$I$_{24}$ (f) MA$_8$Pb$_4$Sn$_4$I$_{24}$ (g) MA$_8$Pb$_3$Sn$_5$I$_{24}$ (h) MA$_8$Pb$_2$Sn$_6$I$_{24}$ (i) MA$_8$PbSn$_7$I$_{24}$ (j) MA$_8$Sn$_8$I$_{24}$.}
	\label{1}
\end{figure}
\begin{figure}[H]
\centering
	\includegraphics[width=15cm, height=16cm]{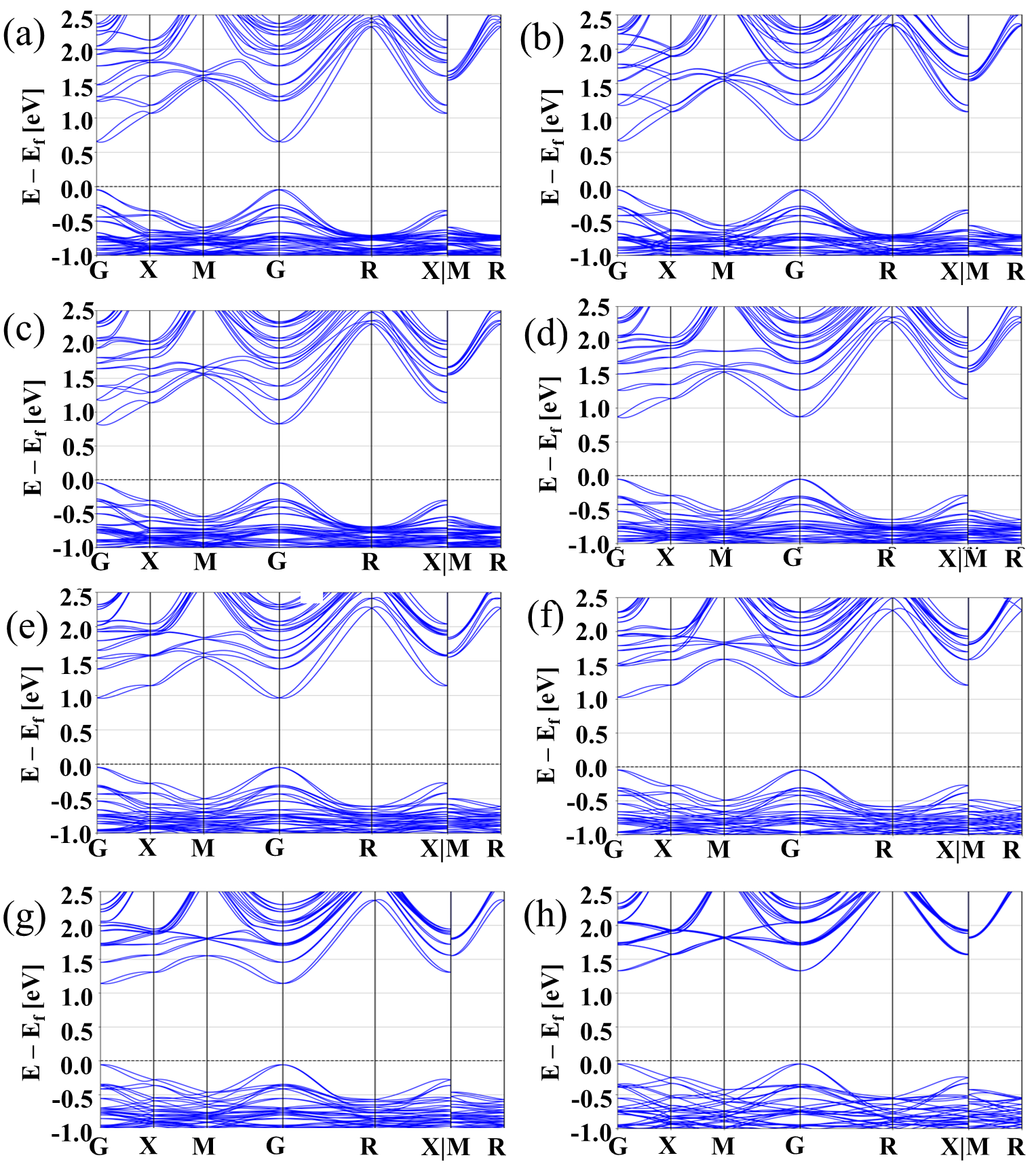}
	\caption{Electronic band structures for completely relaxed configurations calculated using PBE+SOC $\epsilon_{xc}$ functional for (a) MA$_8$Pb$_7$GeI$_{24}$ (b) MA$_8$Pb$_6$Ge$_2$I$_{24}$ (c) MA$_8$Pb$_5$Ge$_3$I$_{24}$ (d) MA$_8$Pb$_4$Ge$_4$I$_{24}$ (e) MA$_8$Pb$_3$Ge$_5$I$_{24}$ (f) MA$_8$Pb$_2$Ge$_6$I$_{24}$ (g) MA$_8$PbGe$_7$I$_{24}$ (h) MA$_8$Ge$_8$I$_{24}$.}
	\label{1}
\end{figure}
\begin{figure}[H]
\centering
	\includegraphics[width=15cm, height=16cm]{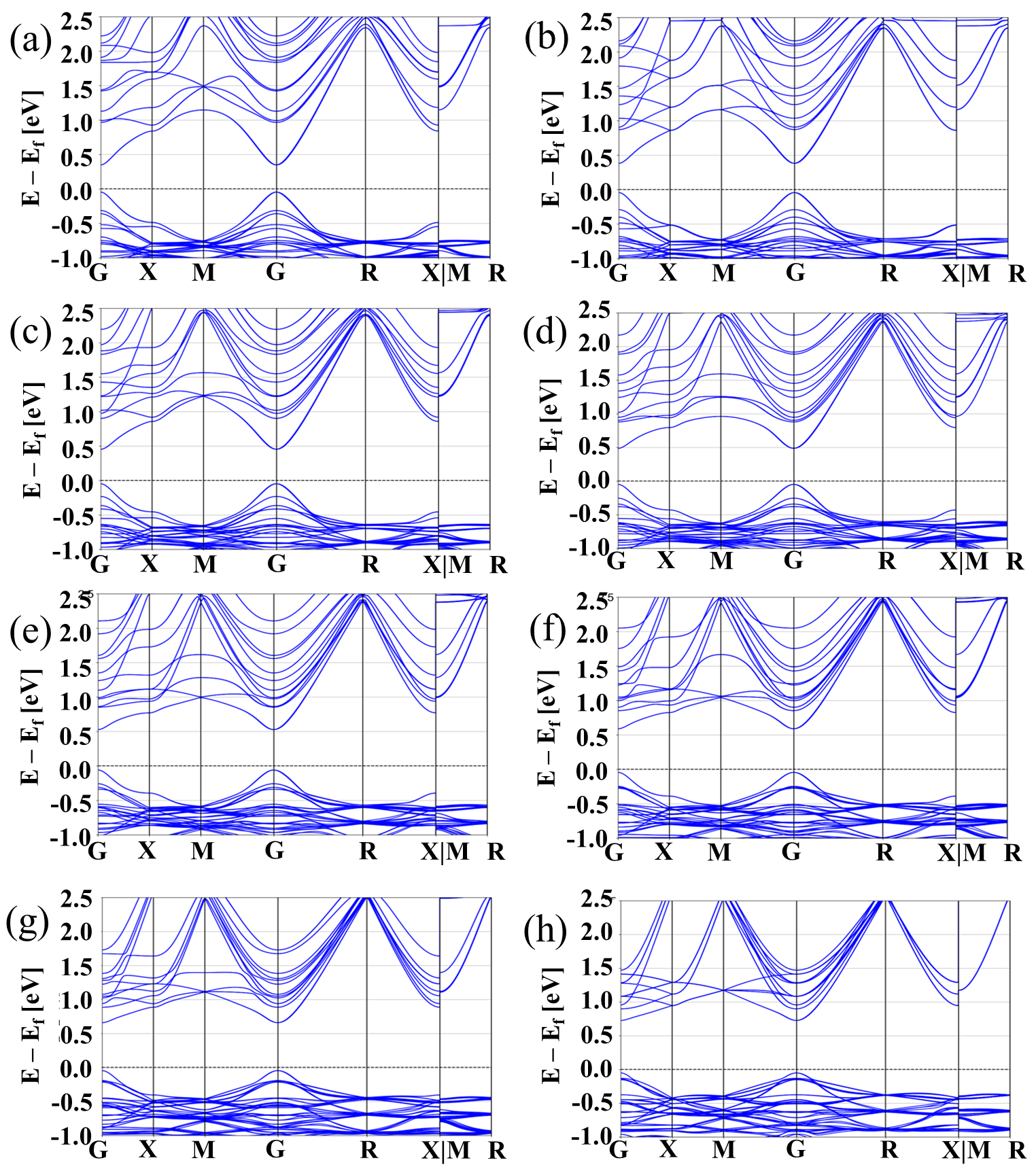}
	\caption{Electronic band structures for MA relaxed configurations calculated using PBE+SOC $\epsilon_{xc}$ functional for (a) MA$_8$Pb$_7$GeI$_{24}$ (b) MA$_8$Pb$_6$Ge$_2$I$_{24}$ (c) MA$_8$Pb$_5$Ge$_3$I$_{24}$ (d) MA$_8$Pb$_4$Ge$_4$I$_{24}$ (e) MA$_8$Pb$_3$Ge$_5$I$_{24}$ (f) MA$_8$Pb$_2$Ge$_6$I$_{24}$ (g) MA$_8$PbGe$_7$I$_{24}$ (h) MA$_8$Ge$_8$I$_{24}$.}
	\label{1}
\end{figure}
\begin{figure}[H]
\centering
	\includegraphics[width=15cm, height=16cm]{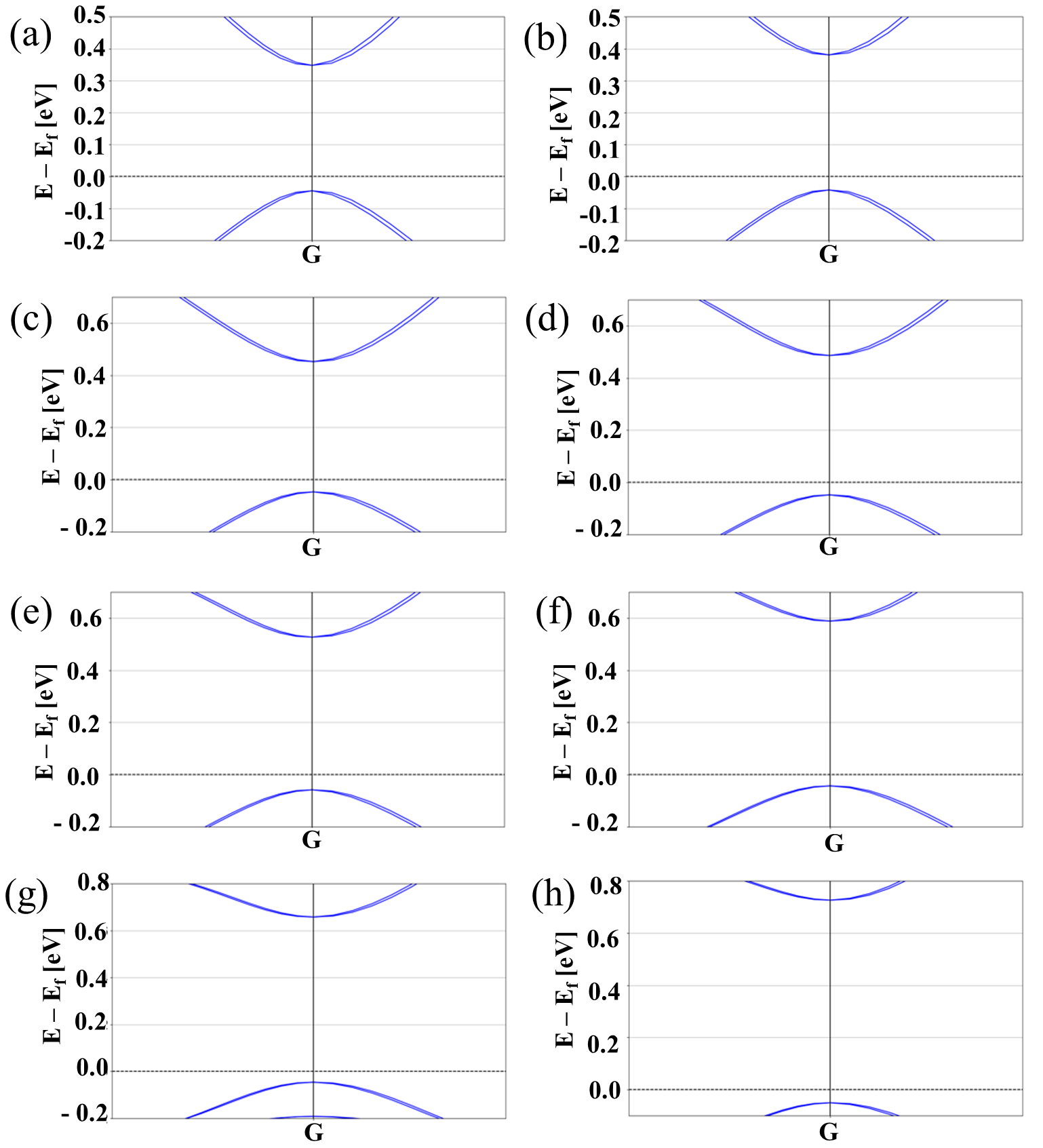}
	\caption{Enlarged view of electronic band structures for MA relaxed configurations around the high-symmetry point G calculated using PBE+SOC $\epsilon_{xc}$ functional for (a) MA$_8$Pb$_7$GeI$_{24}$ (b) MA$_8$Pb$_6$Ge$_2$I$_{24}$ (c) MA$_8$Pb$_5$Ge$_3$I$_{24}$ (d) MA$_8$Pb$_4$Ge$_4$I$_{24}$ (e) MA$_8$Pb$_3$Ge$_5$I$_{24}$ (f) MA$_8$Pb$_2$Ge$_6$I$_{24}$ (g) MA$_8$PbGe$_7$I$_{24}$ (h) MA$_8$Ge$_8$I$_{24}$.}
	\label{1}
\end{figure}

\newpage

\section{pDOS of MA$_8$Pb$_{8-x}$\{Sn/Ge\}$_x$I$_{24}$ perovskites }
\begin{figure}[H]
\centering
	\includegraphics[width=1.0\columnwidth,clip]{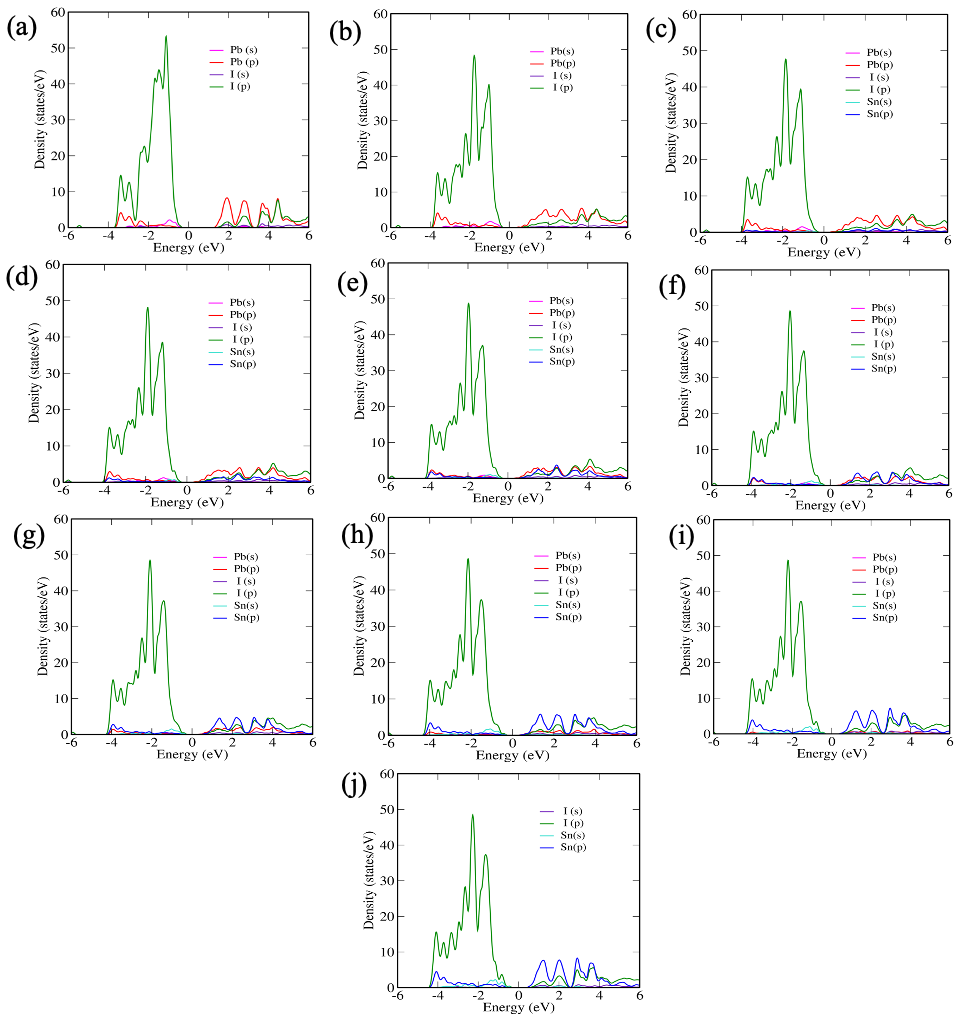}
	\caption{pDOS for completely relaxed configurations calculated using PBE $\epsilon_{xc}$ functional for (a) MA$_8$Pb$_8$I$_{24}$ and using PBE+SOC $\epsilon_{xc}$ functional for (b) MA$_8$Pb$_8$I$_{24}$ (c) MA$_8$Pb$_7$SnI$_{24}$ (d) MA$_8$Pb$_6$Sn$_2$I$_{24}$ (e) MA$_8$Pb$_5$Sn$_3$I$_{24}$ (f) MA$_8$Pb$_4$Sn$_4$I$_{24}$ (g) MA$_8$Pb$_3$Sn$_5$I$_{24}$ (h) MA$_8$Pb$_2$Sn$_6$I$_{24}$ (i) MA$_8$PbSn$_7$I$_{24}$ (j) MA$_8$Sn$_8$I$_{24}$. The Fermi energy is set to zero in the energy axis.}
	\label{1}
\end{figure}
\begin{figure}[H]
\centering
	\includegraphics[width=1.0\columnwidth,clip]{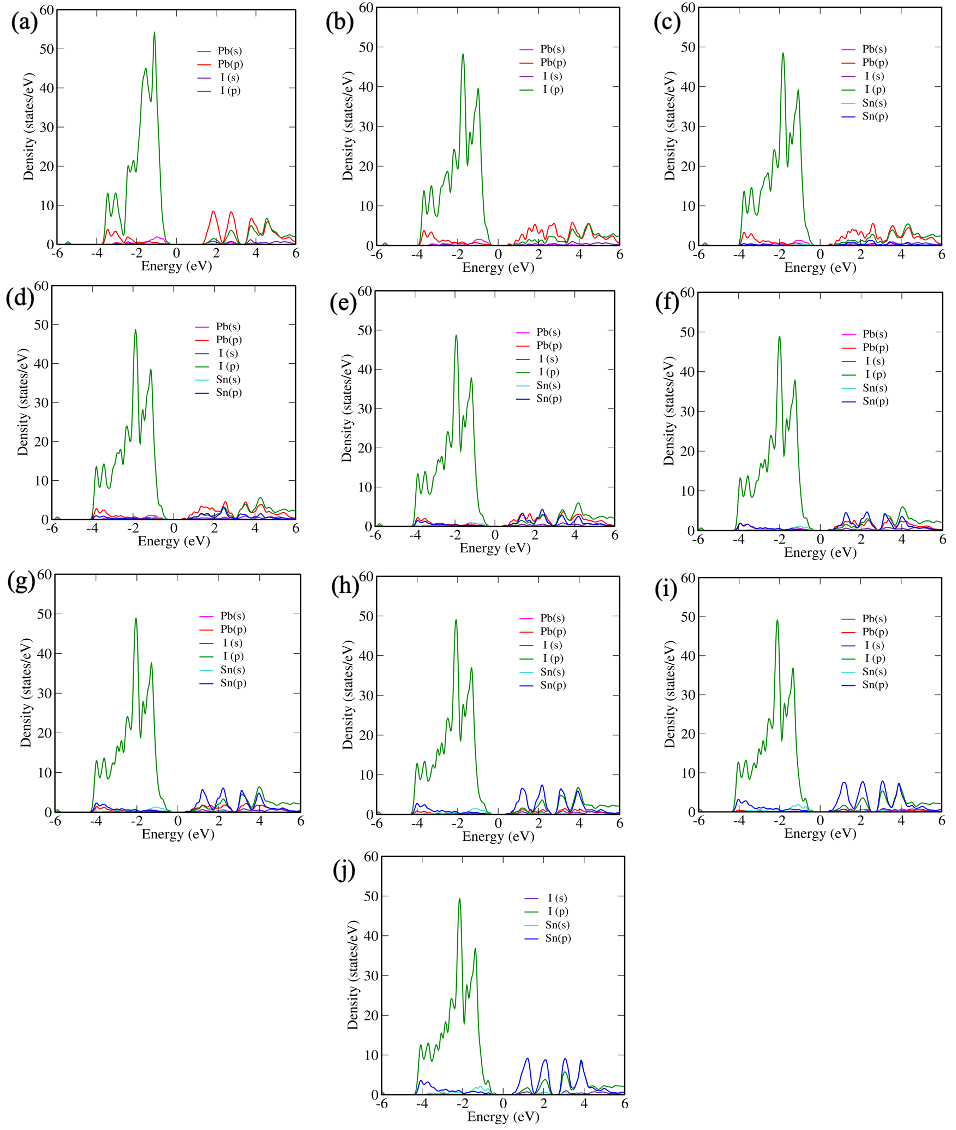}
	\caption{pDOS for only MA relaxed configurations calculated using PBE $\epsilon_{xc}$ functional for (a) MA$_8$Pb$_8$I$_{24}$ and using PBE+SOC $\epsilon_{xc}$ functional for (b) MA$_8$Pb$_8$I$_{24}$ (c) MA$_8$Pb$_7$SnI$_{24}$ (d) MA$_8$Pb$_6$Sn$_2$I$_{24}$ (e) MA$_8$Pb$_5$Sn$_3$I$_{24}$ (f) MA$_8$Pb$_4$Sn$_4$I$_{24}$ (g) MA$_8$Pb$_3$Sn$_5$I$_{24}$ (h) MA$_8$Pb$_2$Sn$_6$I$_{24}$ (i) MA$_8$PbSn$_7$I$_{24}$ (j) MA$_8$Sn$_8$I$_{24}$. The Fermi energy is set to zero in the energy axis.}
	\label{1}
\end{figure}
\begin{figure}[H]
	\includegraphics[width=1.0\columnwidth,clip]{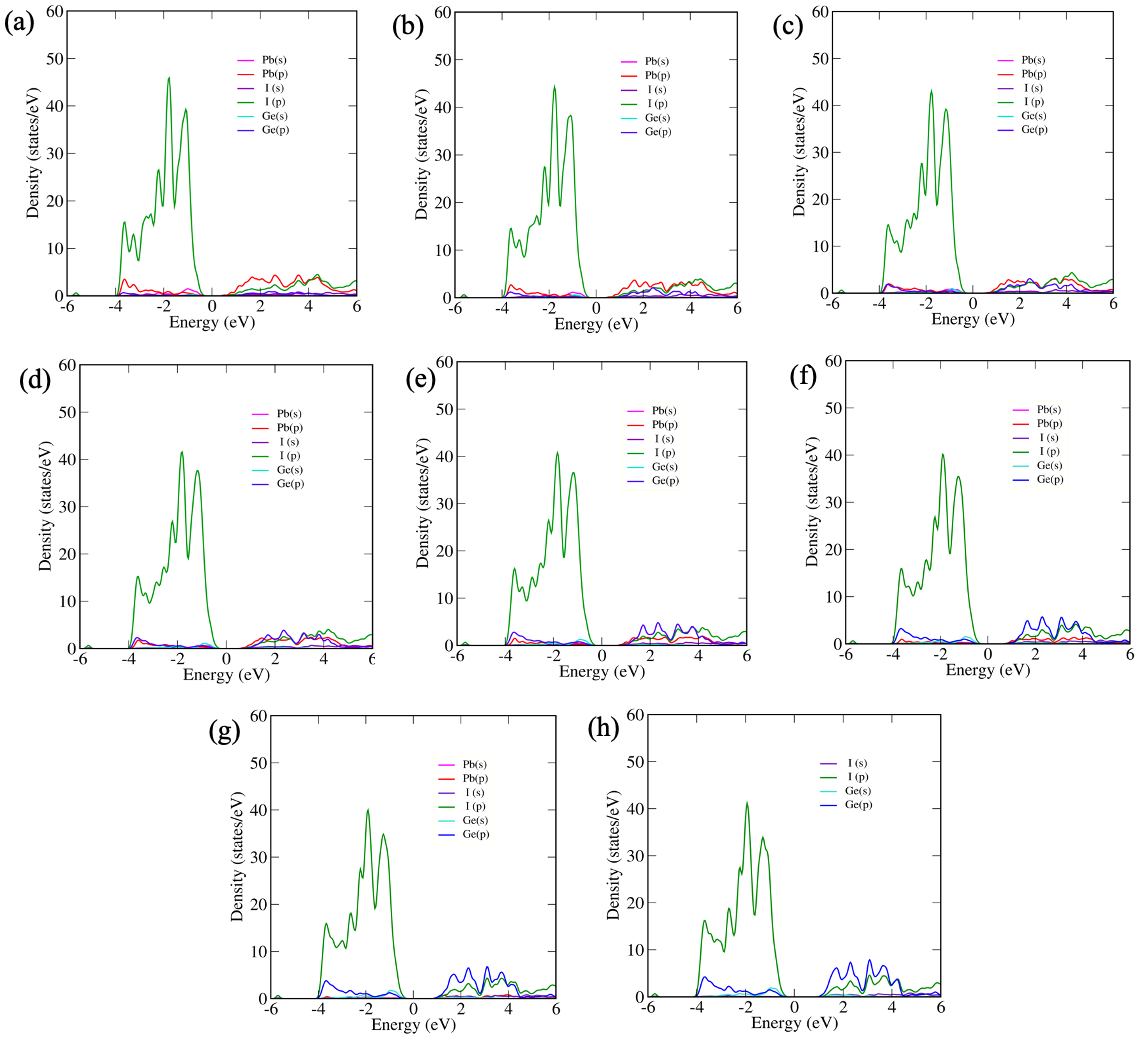}
	\caption{pDOS for completely relaxed configurations calculated using PBE+SOC $\epsilon_{xc}$ functional for (a) MA$_8$Pb$_7$GeI$_{24}$ (b) MA$_8$Pb$_6$Ge$_2$I$_{24}$ (c) MA$_8$Pb$_5$Ge$_3$I$_{24}$ (d) MA$_8$Pb$_4$Ge$_4$I$_{24}$ (e) MA$_8$Pb$_3$Ge$_5$I$_{24}$ (f) MA$_8$Pb$_2$Ge$_6$I$_{24}$ (g) MA$_8$PbGe$_7$I$_{24}$ (h) MA$_8$Ge$_8$I$_{24}$. The Fermi energy is set to zero in the energy axis.}
	\label{1}
\end{figure}
\begin{figure}[H]
	\includegraphics[width=1.0\columnwidth,clip]{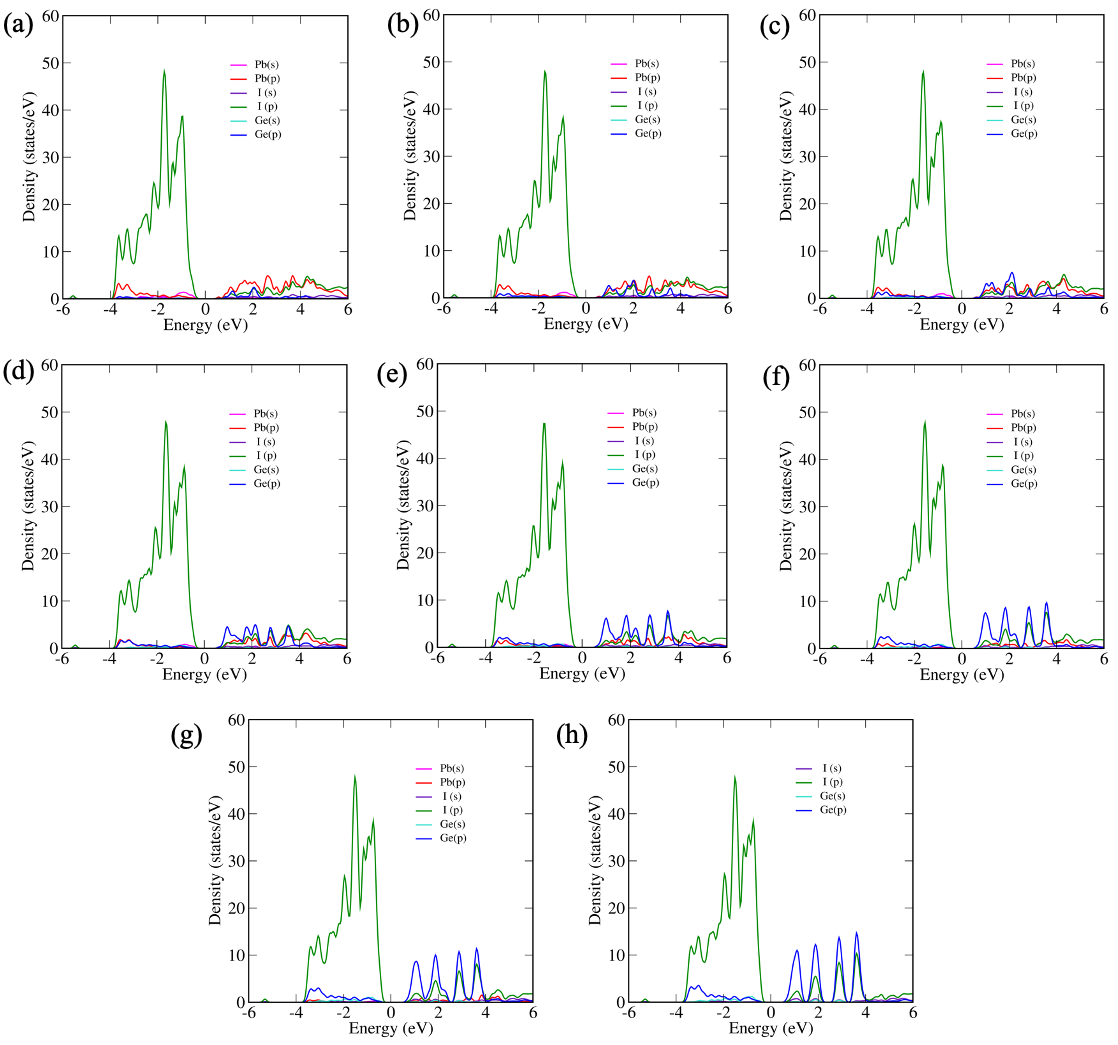}
	\caption{pDOS for only MA relaxed configurations calculated using PBE+SOC $\epsilon_{xc}$ functional for (a) MA$_8$Pb$_7$GeI$_{24}$ (b) MA$_8$Pb$_6$Ge$_2$I$_{24}$ (c) MA$_8$Pb$_5$Ge$_3$I$_{24}$ (d) MA$_8$Pb$_4$Ge$_4$I$_{24}$ (e) MA$_8$Pb$_3$Ge$_5$I$_{24}$ (f) MA$_8$Pb$_2$Ge$_6$I$_{24}$ (g) MA$_8$PbGe$_7$I$_{24}$ (h) MA$_8$Ge$_8$I$_{24}$. The Fermi energy is set to zero in the energy axis.}
	\label{1}
\end{figure}
\newpage
\section{Rashba coefficients of MA$_8$Pb$_{8-x}$\{Sn/Ge\}$_x$I$_{24}$ perovskites}
\begin{table}[H]
	\caption {Rashba coefficients for lower conduction bands in completely relaxed structures}
	\begin{center}
		\begin{adjustbox}{width=0.6\textwidth} 
			\setlength\extrarowheight{+4pt}
			\begin{tabular}[c]{|c|c|c|}  \hline		
				\textbf{Configurations} & \textbf{X–G Direction} & \textbf{G–M Direction}  \\ \hline	
				MA$_8$Pb$_8$I$_{24}$      &  	0.99    &   0.96   \\ \hline
				MA$_8$Pb$_7$SnI$_{24}$     & 	1.23  &   0.83    \\ \hline
				MA$_8$Pb$_6$Sn$_2$I$_{24}$     & 1.03  &  0.84       \\ \hline
				MA$_8$Pb$_5$Sn$_3$I$_{24}$     & 0.69  &   1.04      \\ \hline
				MA$_8$Pb$_4$Sn$_4$I$_{24}$    &  0.76  &   0.85       \\ \hline
				MA$_8$Pb$_3$Sn$_5$I$_{24}$     & 0.59 &    0.89      \\ \hline
				MA$_8$Pb$_2$Sn$_6$I$_{24}$     & 0.42 &   0.83       \\ \hline
				MA$_8$PbSn$_7$I$_{24}$     &  0.14  &    0.81      \\ \hline
				MA$_8$Sn$_8$I$_{24}$     &  0.16 &  0.65          \\ \hline
				MA$_8$Pb$_7$GeI$_{24}$     &  1.11  &  0.80       \\ \hline
				MA$_8$Pb$_6$Ge$_2$I$_{24}$     &  1.07  &  0.65        \\ \hline
				MA$_8$Pb$_5$Ge$_3$I$_{24}$     &  1.36  &  0.69       \\ \hline
				MA$_8$Pb$_4$Ge$_4$I$_{24}$     & 1.07  &  0.43      \\ \hline
				MA$_8$Pb$_3$Ge$_5$I$_{24}$      &  0.56  &   0.78      \\ \hline
				MA$_8$Pb$_2$Ge$_6$I$_{24}$    &  0.56  &    0.34     \\ \hline
				MA$_8$PbGe$_7$I$_{24}$     &  0.48  &    0.37     \\ \hline
				MA$_8$Ge$_8$I$_{24}$     &  0.14  &    0.18      \\ \hline
				
			\end{tabular}
		\end{adjustbox}
		\label{T3}
	\end{center}
\end{table}
\begin{table}[H]
	\caption {Rashba coefficients for lower conduction bands in MA relaxed structures}
	\begin{center}
		\begin{adjustbox}{width=0.6\textwidth} 
			\setlength\extrarowheight{+4pt}
			\begin{tabular}[c]{|c|c|c|}  \hline		
				\textbf{Configurations} & \textbf{X–G Direction} & \textbf{G–M Direction}  \\ \hline	
				MA$_8$Pb$_8$I$_{24}$      &  	No Splitting    &   0.21   \\ \hline
				MA$_8$Pb$_7$SnI$_{24}$     & 	0.01  &   0.23    \\ \hline
				MA$_8$Pb$_6$Sn$_2$I$_{24}$     & 0.02  &  0.24       \\ \hline
				MA$_8$Pb$_5$Sn$_3$I$_{24}$     & 0.03  &   0.27     \\ \hline
				MA$_8$Pb$_4$Sn$_4$I$_{24}$    &  0.03  &   0.29      \\ \hline
				MA$_8$Pb$_3$Sn$_5$I$_{24}$     & 0.04 &    0.34     \\ \hline
				MA$_8$Pb$_2$Sn$_6$I$_{24}$     & 0.05 &  0.32       \\ \hline
				MA$_8$PbSn$_7$I$_{24}$     & 0.01  &    0.39    \\ \hline
				MA$_8$Sn$_8$I$_{24}$     &  0.04 &  0.29        \\ \hline
				MA$_8$Pb$_7$GeI$_{24}$     &  No Splitting  &  0.17       \\ \hline
				MA$_8$Pb$_6$Ge$_2$I$_{24}$     & No Splitting  &  0.14       \\ \hline
				MA$_8$Pb$_5$Ge$_3$I$_{24}$     & No Splitting  &  0.11      \\ \hline
				MA$_8$Pb$_4$Ge$_4$I$_{24}$     & No Splitting  &  0.11     \\ \hline
				MA$_8$Pb$_3$Ge$_5$I$_{24}$      &  No Splitting  &   0.08     \\ \hline
				MA$_8$Pb$_2$Ge$_6$I$_{24}$    &  No Splitting  &   0.07    \\ \hline
				MA$_8$PbGe$_7$I$_{24}$     &  No Splitting  &    0.06   \\ \hline
				MA$_8$Ge$_8$I$_{24}$     &  No Splitting  &   0.10     \\ \hline
				
			\end{tabular}
		\end{adjustbox}
		\label{T3}
	\end{center}
\end{table}

\newpage

\section{Variation of Rashba parameters of MA$_8$Pb$_{8-x}$\{Sn/Ge\}$_x$I$_{24}$ perovskites on application of axial strain}
\begin{figure}[H]
	\includegraphics[width=18cm]{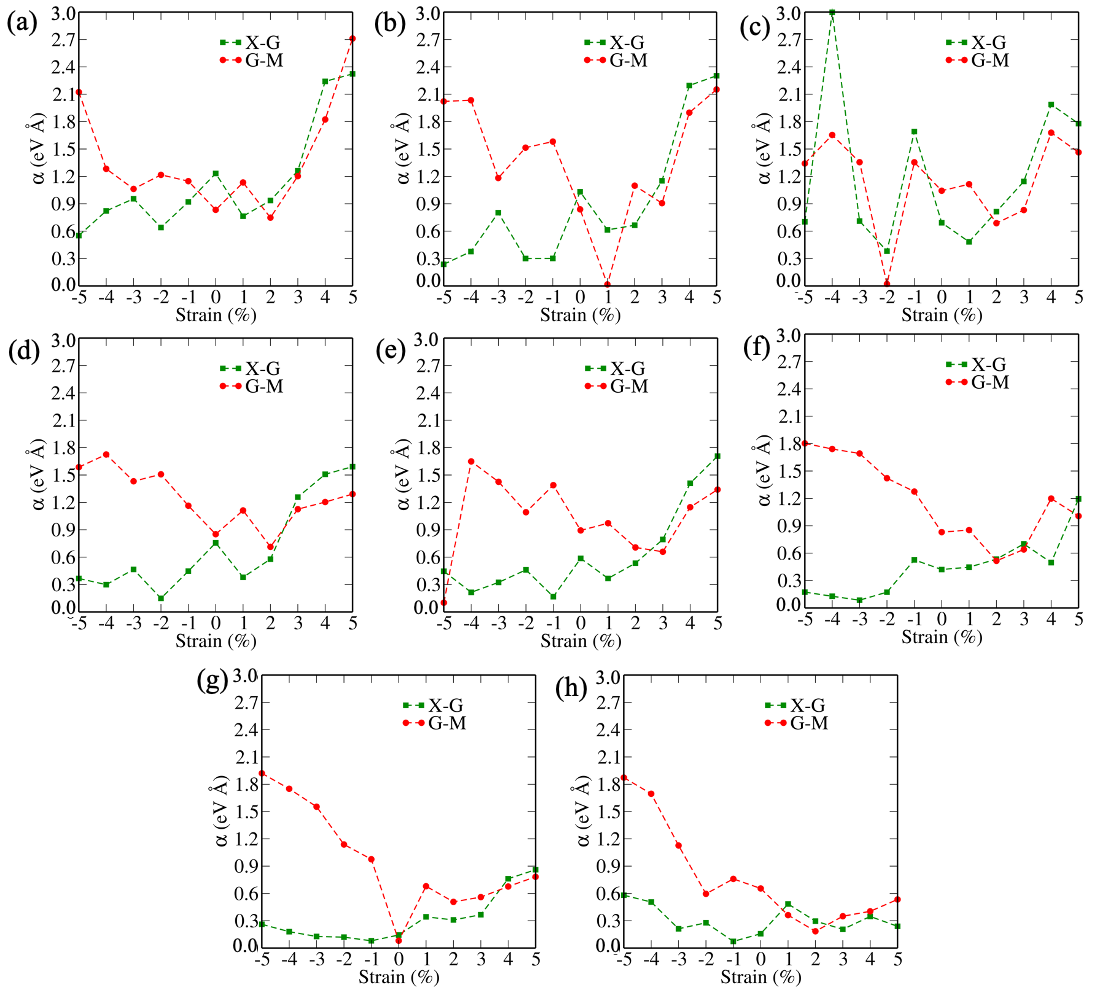}
	\caption{Rashba coefficients for lower conduction bands when strain is applied along ‘\emph{c}’ axis in (a) MA$_8$Pb$_7$SnI$_{24}$ (b) MA$_8$Pb$_6$Sn$_2$I$_{24}$ (c) MA$_8$Pb$_5$Sn$_3$I$_{24}$ (d) MA$_8$Pb$_4$Sn$_4$I$_{24}$ (e) MA$_8$Pb$_3$Sn$_5$I$_{24}$ (f) MA$_8$Pb$_2$Sn$_6$I$_{24}$ (g) MA$_8$PbSn$_7$I$_{24}$ (h) MA$_8$Sn$_8$I$_ {24}$}
	\label{1}
\end{figure}
\begin{figure}[H]
	\includegraphics[width=18cm]{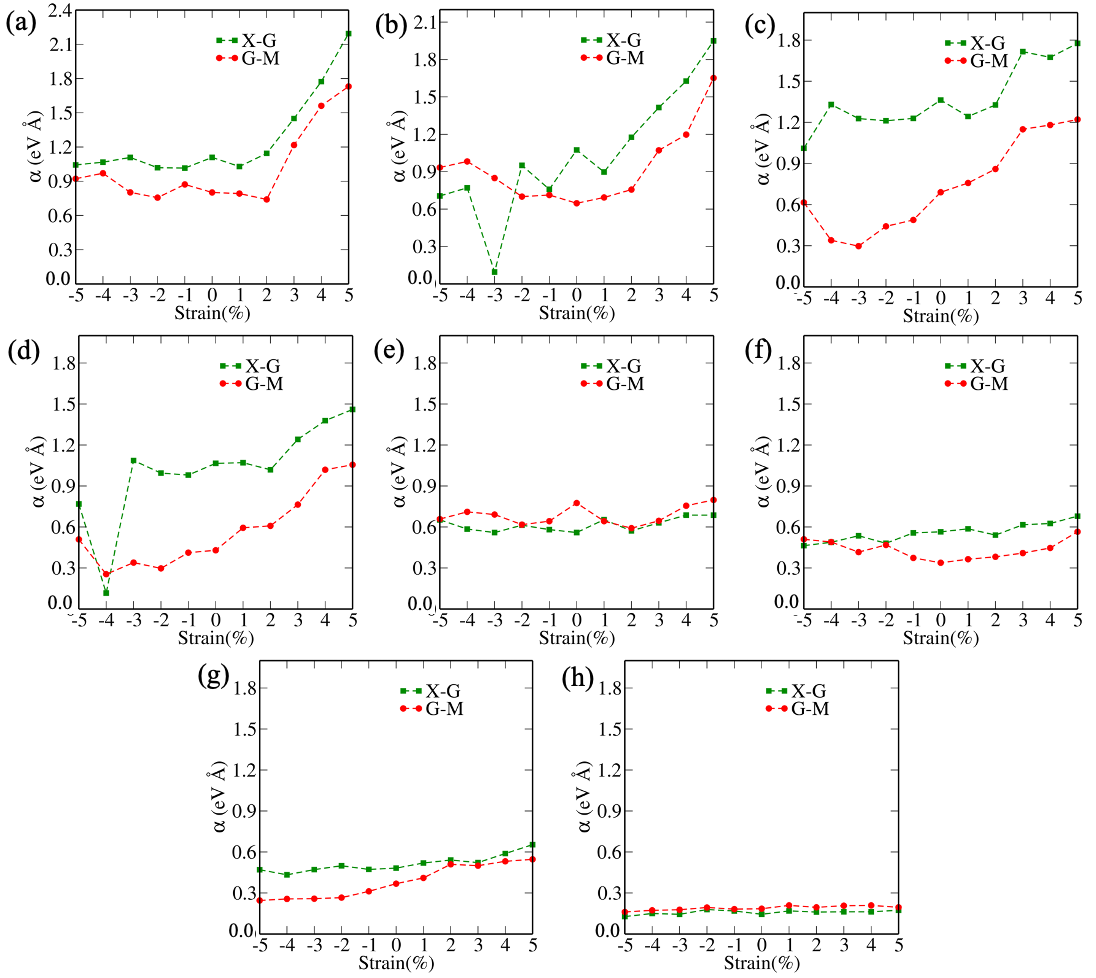}
	\caption{Rashba coefficients for lower conduction bands when strain is applied along ‘\emph{c}’ axis in (a) MA$_8$Pb$_7$GeI$_{24}$ (b) MA$_8$Pb$_6$Ge$_2$I$_{24}$ (c) MA$_8$Pb$_5$Ge$_3$I$_{24}$ (d) MA$_8$Pb$_4$Ge$_4$I$_{24}$ (e) MA$_8$Pb$_3$Ge$_5$I$_{24}$ (f) MA$_8$Pb$_2$Ge$_6$I$_{24}$ (g) MA$_8$PbGe$_7$I$_{24}$ (h) MA$_8$Ge$_8$I$_ {24}$}
	\label{1}
\end{figure}
\newpage
\section{Mechanical properties of MA$_8$Pb$_{8-x}$\{Sn/Ge\}$_x$I$_{24}$ perovskites}

The various mechanical properties, i.e., bulk moduli (\textit{B}), shear moduli (\textit{G}), Pugh’s ratio (\textit{B/G}) and Poisson’s ratio ($\nu$) can be calculated from elastic constants \textit{C$_\textit{{11}}$}, \textit{C$_\textit{{12}}$} and \textit{C$_\textit{{44}}$} by using the following equations\cite{hill1952elastic,roknuzzaman2017towards}: 
\begin{equation}
	\begin{aligned}
		B = B_V = B_R = (C_\textit{{11}} + 2C_\textit{{12}}) / 3
		\label{eq1}
	\end{aligned}
\end{equation}
\begin{equation}
	\begin{aligned}
		G_V = (C_\textit{{11}} - C_\textit{{12}} + 3C_\textit{{44}}) / 5
		\label{eq2}
	\end{aligned}
\end{equation}
\begin{equation}
	\begin{aligned}
		G_R = 5 (C_\textit{{11}} - C_\textit{{12}}) C_\textit{{44}} / 4C_\textit{{44}} + 3(C_\textit{{11}} - C_\textit{{12}})
		\label{eq3}
	\end{aligned}
\end{equation}
\begin{equation}
	\begin{aligned}
		G = (G_V + G_R) / 2
		\label{eq4}
	\end{aligned}
\end{equation}
\begin{equation}
	\begin{aligned}
		\nu = (3B - 2G) / (6B + 2G)
		\label{eq5}
	\end{aligned}
\end{equation}

In case of cubic phase, following Born's stability criteria\cite{born1940stability} can be used to evaluate the mechanical stability: 
\begin{equation}
	\begin{aligned}
		C_\textit{{11}} > 0, C_\textit{{44}} > 0, C_\textit{{11}} > |C_\textit{{12}}|, (C_\textit{{11}} + 2C_\textit{{12}}) > 0
	\end{aligned}
\end{equation}

Further, Pugh's ratio and Poisson's ratio have 1.75 and 0.26, respectively as the critical values, i.e., if  Pugh's ratio (Poisson's ratio) is greater than 1.75 (0.26), the material is ductile and if it is less than 1.75 (0.26), the material is brittle.\cite{pugh1954xcii,hadi2017elastic}

\begin{table}[H]
	\caption { The calculated elastic constants \textit{C}${ij}$ (GPa), bulk modulus \textit{B} (GPa), shear modulus \textit{G} (GPa), Pugh's ratio (\textit{B/G}), and Poisson's ratio $\nu$ of mechanically stable configurations.}
	\begin{center}
		\begin{adjustbox}{width=0.9\textwidth} 
			\setlength\extrarowheight{+4pt}
			\begin{tabular}[c]{|c|c|c|c|c|c|c|c|} \hline		
				\textbf{Configurations} & \textbf{\textit{C$_{\textit{11}}$}} & \textbf{\textit{C$_{\textit{12}}$}} & \textbf{\textit{C$_{\textit{44}}$}} & \textbf{\textit{B}} & \textbf{\textit{G}}  & \textbf{\textit{B/G}}   &   \textbf{$\nu$}     \\ \hline
				MA$_8$Pb$_8$I$_{24}$      &  365.57  &123.88 & 24.06 & 204.44    &        49.08   &   4.17   &  0.39  \\ \hline
				MA$_8$Pb$_6$Sn$_2$I$_{24}$      &  148.97  & 18.99  & 36.39 & 62.32     &   46.00      &   1.35   &  0.20  \\ \hline
				MA$_8$Pb$_5$Sn$_3$I$_{24}$        &  127.93  & -17.90 & 25.40 & 30.71     &   34.35       &    0.89   &  0.09  \\ \hline
				MA$_8$Pb$_4$Sn$_4$I$_{24}$       & 212.17  &  34.72  & 26.51 & 93.87     &       44.12  &   2.13      &  0.30 \\ \hline
				MA$_8$Pb$_2$Sn$_6$I$_{24}$      &  212.40 & 71.79 & 31.60 & 118.66     &        43.80     &   2.71    & 0.34  \\ \hline
				MA$_8$Pb$_6$Ge$_2$I$_{24}$       &  134.81  & 58.78  & 37.80 & 84.13     &        37.89     &    2.22   &  0.30    \\ \hline
				MA$_8$Pb$_5$Ge$_3$I$_{24}$       &  91.77   & 64.69 &26.90 & 73.72
	            &       20.42     &   3.61   &  0.37   \\ \hline
				MA$_8$Pb$_2$Ge$_6$I$_{24}$       &  47.51  &  21.36  & 27.41 & 30.07   &     20.37   &   1.48    &   0.22   \\ \hline
				
			\end{tabular}
		\end{adjustbox}
		\label{T3}
	\end{center}
\end{table}

\newpage

\providecommand{\latin}[1]{#1}
\makeatletter
\providecommand{\doi}
{\begingroup\let\do\@makeother\dospecials
	\catcode`\{=1 \catcode`\}=2 \doi@aux}
\providecommand{\doi@aux}[1]{\endgroup\texttt{#1}}
\makeatother
\providecommand*\mcitethebibliography{\thebibliography}
\csname @ifundefined\endcsname{endmcitethebibliography}
{\let\endmcitethebibliography\endthebibliography}{}
